\documentclass[12pt, titlepage]{utarticle}
\usepackage{amssymb, amsmath, amsopn, amsthm,amscd,amsfonts}
\usepackage{graphicx}
\usepackage{hyperref} 

\numberwithin{equation}{section}

\newcommand\MM{\mathcal{M}}

\begin{document}
\preprint{UTTG-09-07 \\
}

\title{New Supergravity Backgrounds Dual to ${\mathcal N}=1$\\[.4cm]
    SQCD-like Theories with $N_f=2N_c$}

\author{Elena C{\'a}ceres,
	 \address{
	  Facultad de Ciencias,\\
      Universidad de Colima, \\
      Bernal D{\'i}az del Castillo 340,\\
      C.P. 28045 Colima, Colima, M{\'e}xico\\
     }
	Raphael Flauger, Matthias Ihl and Timm Wrase
	 \address{
	  Theory Group, Department of Physics,\\
      University of Texas,\\
      Austin, TX 78712, USA\\
      {~}\\
      {\rm Email}\\
      \emailt{elenac@ucol.mx}\\ 
      \emailt{flauger@physics.utexas.edu}\\
      \emailt{msihl@zippy.ph.utexas.edu}\\
      \emailt{wrase@zippy.ph.utexas.edu}}
	}

\Abstract{We present new supergravity backgrounds generated by $N_c$ D5-branes, wrapping the $S^2$ of the resolved conifold, in the presence of $N_f = 2 N_c$ smeared flavor D5-branes. The smearing allows us to take their backreaction on the geometry into account. We discuss the consistency, stability, and supersymmetry of these types of setups. We find near horizon geometries that we expect to be supergravity duals of SQCD-like theories with $N_f= 2N_c$. From these backgrounds we numerically extract rectangular Wilson loops and beta functions of the dual field theory for the regime where our approximations are valid.}

\maketitle

\tableofcontents

\newpage
\section{Introduction}
The gauge/gravity correspondence \cite{Maldacena:1997re},\cite{Witten:1998qj} (for a review and additional references, see \cite{Aharony:1999ti}) is  undoubtedly a powerful tool to study strongly coupled gauge theories. Within this framework, the Chamseddine-Volkov-Maldacena-Nu\~nez (CVMN) \cite{Chamseddine:1997nm},\cite{Chamseddine:1997mc},\cite{Maldacena:2000yy} and Klebanov-Strassler \cite{Klebanov:2000hb} backgrounds are  landmarks in our understanding
of $\mathcal N =1$ Super-Yang-Mills gravity duals.  These backgrounds have been thoroughly studied and extended in many ways in an attempt to obtain more realistic models. In particular, since they do not contain fundamental matter, a natural question that arises is how to include dynamical quarks. The flavor degrees of freedom
imply the addition of an open string sector. This is achieved by adding  flavor branes. In the probe approximation \cite{Karch:2002sh} a small number of flavor
branes is added to the background and a new, tunable, scale $m_q$ appears in the problem leading to  interesting effects in the predicted  meson
spectra,  the understanding of chiral symmetry breaking and phase transition \cite{Kruczenski:2003be}.
But in order to neglect the backreaction of the flavor branes we are bound to take $N_f/N_c \rightarrow 0$,  in the large $N_c$ limit, and thus, some of the SQCD physics will not be captured by this approximation.
Recently, Casero, Nu\~nez and Paredes \cite{Casero:2006pt} proposed a procedure to add a large number of flavors, $N_f/N_c \sim 1$, to the CVMN background. The idea is to introduce $N_f$ flavor branes that extend along the same $3+1$ gauge theory dimensions as the $N_c$ color branes that generated the background, wrap an appropriate cycle and are smeared along the other angular directions.\footnote{The same technique has been used to embed a large number of flavor branes into the Klebanov-Witten~\cite{Benini:2006hh} and Klebanov-Strassler \cite{Benini:2007gx} backgrounds.}
In \cite{Casero:2006pt} the authors proposed this setup,  solved the BPS equations asymptotically (for  small and large values of the radial variable) and studied several gauge theory aspects of the solutions. This first step in the quest of a SQCD  dual has opened a plethora of  issues   to be studied, tested and understood.

In the present work  we find new families of supergravity solutions  for the particular case of $N_f= 2N_c$. We consider flavor branes extending  along all non-compact directions and wrapping  a trivial cycle in the compact space. We remark on the consistency, study the stability and supersymmetry of this embedding, and  present several families of solutions. We find two two-parameter families of solutions and are able to write analytic expressions for them in certain regions. We also find a one-parameter family of solutions that we can only study numerically.  The $N_f=2N_ c$ solution presented in Appendix D of \cite{Casero:2006pt} is included in our solutions.
Our study of the   $N_f=2N_ c$  solutions is based on solving BPS equations. As such, all solutions presented are supergravity solutions but not all of them are dual to gauge theories. The near-horizon geometries that are dual to  SQCD-like theories are contained within our solutions. We identify them as the above-mentioned one-parameter family  and calculate the Wilson loops and beta functions for these backgrounds.

The paper is organized as follows: In Section \ref{sec:review} we briefly review the CVMN background, a supergravity background dual to ${\mathcal N}=1$ SYM in four dimensions. We then  summarize the  procedure proposed in \cite{Casero:2006pt} to include a large number of smeared flavor D5-branes into this background taking into account their backreaction on the geometry. In Section \ref{sec:sols} we  examine the validity of this approach and discuss some subtleties. We then describe in detail a new set of solutions to the BPS equations for $N_f=2N_c$.  In Section \ref{sec:sqcd} we study  field theory quantities; we calculate the  Wilson loops and beta functions of these solutions. We conclude in Section \ref{sec:concl}, pointing out some issues and extensions that deserve further study. In  Appendix~\ref{ap:stability} we analyze the  stability   of a probe  D5-brane in the CVMN  and in our backreacted background.  In Appendix~\ref{ap:BPS-eqs} we derive the BPS equations for our particular setup following \cite{Nunez:2003cf} and show that a single probe brane in this backreacted background is kappa symmetric. In Appendix~\ref{ap:solutions} we present the derivation of the analytic solutions.

\section{Review of Previous Work}\label{sec:review}
In this section we review the CVMN background as well as how to embed a large number of flavor branes into this background following \cite{Casero:2006pt}.
The CVMN background was found by Chamseddine and Volkov~\cite{Chamseddine:1997nm},\cite{Chamseddine:1997mc} and later interpreted by Maldacena and Nu\~nez~\cite{Maldacena:2000yy} as the near horizon geometry of a large number of NS5- or D5-branes wrapping an $S^2$. Modulo the usual subtleties involving Kaluza-Klein modes it is dual to pure ${\mathcal N}=1$ SYM in four dimensions~\cite{Maldacena:2000yy}. 
The construction in \cite{Casero:2006pt} allows the embedding of a large number of flavor branes into this background. This leads to a new supergravity background conjectured to be dual to ${\mathcal N}=1$ SQCD with a large number of flavors.

\subsection{The Chamseddine-Volkov-Maldacena-Nu\~nez (CVMN) Background}
In two nice papers Chamseddine and Volkov found a monopole solution to $SU(2)\times SU(2)$ gauged supergravity in four dimensions \cite{Chamseddine:1997nm} and lifted it to a solution of ten-dimensional supergravity \cite{Chamseddine:1997mc}. The significance of this solution was not understood until Maldacena and Nu\~nez rederived it  as the near horizon geometry of NS5-branes whose worldvolume was compactified on an $S^2$ and noticed that as such it should be dual to pure ${\mathcal N}=1$ SYM in four dimensions at low energies \cite{Maldacena:2000yy}.\footnote{This was further supported by the more detailed analysis in \cite{Papadopoulos:2000gj},\cite{Gubser:2001eg}}\\ This can be understood as follows. As is well known, in the UV, the correct description of parallel NS5-branes is given by little string theory (LST) \cite{Seiberg:1997zk},\cite{Losev:1997hx} (for a brief review and additional references, see \cite{Aharony:1999ks}). In the IR, the S-dual picture involving parallel D5-branes is more appropriate \cite{Itzhaki:1998dd}. Their low energy degrees of freedom are described by $(5+1)$-dimensional SYM with sixteen supercharges. If two of the directions of the brane wrap a compact manifold, then, at low energies, the degrees of freedom of the D5-branes are described by a $(3+1)$-dimensional gauge theory. However, if this 2-manifold is curved, in general supersymmetry is not preserved. In the CVMN solution, the 2-manifold is chosen to be a sphere\footnote{Other topologies for the 2-manifolds are possible and are dual to $\mathcal N=1$ SYM with matter in the adjoint representation \cite{Gauntlett:2001ps}.}. So in order to preserve some supersymmetry, the normal bundle of the brane has to be appropriately twisted \cite{Bershadsky:1995qy}. For the twist chosen, four supercharges are preserved and the low energy degrees of freedom of the branes are described by pure ${\mathcal N}=1$ SYM in four dimensions.\footnote{In our discussion we have been somewhat careless about the fields describing the motion of the brane in the transverse directions. These acquire a mass due to the twist and disappear from the spectrum at low energies.}\\ Notice again that this supergravity solution will only be dual to a four-dimensional field theory at low energies ({\it i.e.}, for small values of the radial coordinate). At higher energies, the modes of the gauge theory start to explore the $S^2$ and the theory first becomes six-dimensional ${\mathcal N}=1$ SYM and then, at even higher energies, the blowing-up of the dilaton forces us to S-dualize and a little string theory completes the model in the UV.\\ 
The corresponding supergravity solution, after lifting it up to ten dimensions, has the topology of $\mathbb{R}^{1,3} \times \mathbb{R}^3 \times S^3$.\footnote{At first sight the metric suggests $\mathbb{R}^{1,3} \times \mathbb{R}\times S^2 \times S^3$ topology. A more careful study shows that the singularity at $r=0$ is merely a coordinate singularity and that the solution indeed has $\mathbb{R}^{1,3} \times \mathbb{R}^3 \times S^3$ topology.} The metric in the Einstein frame reads,
\begin{equation}\label{eq:metricansatz}
ds_{10}^2 = \alpha' g_s N_c e^{\phi/2} \left[\vphantom{\frac{e^{2g}}{4}} \frac{1}{\alpha' g_s N_c} dx_{1,3}^2
+ e^{2h} (d\theta^2 + \sin^2 \theta d\varphi^2) + dr^2 + \frac{1}{4} (w^a - A^a)^2 \right],
\end{equation}
where $\phi$ is the dilaton. The angles are spherical polar coordinates $\theta \in [0,\pi]$ and $\varphi \in [0, 2\pi)$ parametrizing a two-sphere. The $w^a$ are the $SU(2)$ left-invariant one-forms on the $S^3$,
\begin{align}\label{eq:leftinv}
w^1 &=\cos \psi d\tilde{\theta} + \sin \psi \sin \tilde{\theta} d \tilde{\varphi},\nonumber\\
w^2 &=-\sin \psi d\tilde{\theta} + \cos \psi \sin \tilde{\theta} d \tilde{\varphi},\\\nonumber
w^3 &=d\psi + \cos \tilde{\theta} d \tilde{\varphi},
\end{align}
where $\tilde \theta, \tilde \varphi, \psi$ are Euler angles on the 3-sphere with conventions chosen such that $0 \leq$~$\tilde{\theta}$~$ \leq$~$\pi$, $0 \leq \tilde{\varphi} < 2 \pi$ and $0 \leq \psi < 4 \pi$. The twisting is achieved by choosing an embedding of the nontrivial $U(1)$ part of the spin connection into the R-symmetry group $SO(4) \sim SU(2)_R \times SU(2)_L$ (the group of isometries of the $S^3$).\footnote{A different embedding is possible leading to a theory with $\mathcal{N}=2$ supersymmetry \cite{Gauntlett:2001ps}.} The connection one-forms $A^a$ ($a=1,2,3$) can be written in terms of a function $a(r)$ and the angles $(\theta,\varphi)$ as follows
\begin{equation}\label{eq:Aa}
A^1=-a(r) d\theta,\quad A^2= a(r)\sin \theta d \varphi, \quad A^3=-\cos \theta d\varphi.
\end{equation}
For the metric ansatz (\ref{eq:metricansatz}), one obtains a regular supersymmetric solution when the functions $a(r), h(r)$ and the dilaton $\phi(r)$ are
\begin{subequations}\label{eq:MNsol}
\begin{align}
a(r) &= \frac{2r}{\sinh 2r},\\
e^{2h(r)}&= r \coth 2r - \frac{r^2}{\sinh^2 2r} -\frac{1}{4},\\
e^{-2\phi(r)} &=e^{-2\phi_0} \frac{2 e^h}{\sinh 2r},
\end{align}
\end{subequations}
where $\phi_0$ is the value of the dilaton at $r=0$. Near the origin $r=0$ the function $e^{2h}$ behaves as $e^{2h} \sim r^2$ and the metric is non-singular. The solution of the type IIB supergravity requires a Ramond-Ramond three-form $F_{(3)}$ given by
\begin{equation}
\frac{1}{\alpha' g_s N_c} F_{(3)} = - \frac{1}{4} (w^1 -A^1) \wedge (w^2 -A^2) \wedge (w^3 -A^3) + \frac{1}{4} \sum_a F^a \wedge (w^a -A^a),
\end{equation}
where $F^a$ is the field strength of the $SU(2)$ gauge field $A^a$, defined as
\begin{equation}
F^a = dA^a + \frac{1}{2} \epsilon_{abc} A^b \wedge A^c.
\end{equation}
The different components of $F^a$ read,
\begin{equation}
F^1 = -a' dr \wedge d\theta, \quad F^2= a' \sin \theta dr \wedge d\varphi, \quad F^3 = (1-a^2) \sin \theta d\theta \wedge d \varphi,
\end{equation}
where the prime denotes the derivative with respect to r. \\
Let us stress that this configuration is non-singular. We should point out that there is a one-parameter family of singular non-abelian solutions that interpolates between this regular non-abelian solution and the singular abelian solution of \cite{Maldacena:2000yy} with $a(r)=0$. This family was first found in \cite{Gubser:2001eg}.\\ 

\subsection{Adding Flavor to the CVMN Background}
A possible way to add flavor to the CVMN background is to consider supersymmetric embeddings of flavor D5-branes that extend along the Minkowski directions, along the radial direction, as well as along a trivial cycle inside the compact directions~\cite{Nunez:2003cf}. The analysis in \cite{Nunez:2003cf} is done in the probe approximation along the lines of the original work by Karch and Katz \cite{Karch:2002sh}, {\it i.e.}, the number of flavors is taken to be much smaller than the number of colors, $N_f \ll N_c$, and thus the backreaction of the flavor D5-branes can be neglected.\\
Casero, Nu{\~n}ez and Paredes \cite{Casero:2006pt}, went one step further and presented solutions that incorporate the backreaction of the flavor D5-branes. The geometries they construct depend on the ratio $x=N_f/N_c$ which can be kept of order one; they are singular at the origin, but the singularity is a "good" one in the sense of the criterion in \cite{Gubser:2000nd} which means that the metric component $g_{tt}$ remains finite in the limit $r \rightarrow 0$. Everywhere else the geometry is smooth and the curvature small as long as $g_sN_c\gg1$ as will be discussed below. These backgrounds are conjectured to be dual to ${\mathcal N}=1$ SQCD in four dimensions with a large number of flavors, up to the same caveats concerning the decoupling of the KK modes that were already present in the discussion of the original CVMN background without flavor.\\
The general strategy is the following \cite{Casero:2006pt}: One introduces a deformation of the CVMN background due to the presence of flavor D5-branes, derives the corresponding BPS equations (see Appendix B of \cite{Casero:2006pt}), and finally attempts to solve them. The flavor D5-branes are taken to extend along the $(x^0,x^1,x^2,x^3,\psi,r)$ directions and are smeared\footnote{The process of smearing will be explained in detail in section \ref{sec:sols}.} over the $(\theta,\varphi,\tilde{\theta},\tilde{\varphi})$ directions. These branes can be shown to preserve the same supersymmetry as the background for arbitrary values of the angles $\theta, \varphi, \tilde{\theta}, \tilde{\varphi}$ \cite{Nunez:2003cf}. Moreover, the smeared flavor branes will be sources for the RR 3-form, resulting in RR fluxes in the deformed background that can be observed as a ``violation" of the original Bianchi identity. \\
To this end, one introduces an ansatz for the deformation of the CVMN background ($H_{(3)}=0,\, F_{(5)}=0$)
\begin{align}\label{eq:defmetansatz}
ds_{10}^2 &= e^{2f(r)} \left[\vphantom{\frac{e^{2g}}{4}}dx_{1,3}^2+dr^2
+ e^{2h(r)} (d\theta^2 + \sin^2 \theta d\varphi^2) + \right.\nonumber\\
& \quad \left. +\frac{e^{2g(r)}}{4} \left( (w^1 +a(r)d\theta)^2 +(w^2 -a(r) \sin \theta d \varphi)^2 \right) + \frac{e^{2k(r)}}{4}(w^3 + \cos \theta d \varphi)^2 \right],
\end{align}
where we have set $\alpha' g_s =1$ and $N_c$ has been absorbed into $e^{2h}, e^{2g},e^{2k}$ and $dr^2$. The dilaton is given by $\phi = 4 f$. The left-invariant one-forms $w^a$ on $S^3$ are the ones given in (\ref{eq:leftinv}). The RR 3-form field strength reads
\begin{align}\label{eq:F3o}
F_{(3)}& = \frac{N_c}{4} \left[ -(w^1 +b(r) d\theta)\wedge (w^2 - b(r) \sin \theta d\varphi) \wedge (w^3 +\cos \theta d \varphi)  \right.\nonumber\\
&\left. + b' dr \wedge (-d \theta \wedge w^1 + \sin \theta d\varphi \wedge w^2) + (1-b(r)^2) \sin \theta d \theta \wedge d\varphi \wedge w^3\right],
\end{align}
and automatically satisfies the Bianchi identity $dF_{(3)}=0$.\\
It is convenient to introduce a basis of vielbeins for the above metric ansatz (\ref{eq:defmetansatz})
\begin{equation}
\begin{array}{ll}\label{eq:vielbein}
e^{x^\mu} = e^f dx^\mu, & e^1 = \frac{1}{2} e^{f+g}(w^1 + a(r) d\theta), \\
e^\rho = e^f dr = e^{f+k} d\rho, \quad & e^2 = \frac{1}{2} e^{f+g}(w^2 - a(r)\sin \theta d\varphi), \\
e^{\theta} = e^{f+h} d\theta, & e^3 = \frac{1}{2} e^{f+k}(w^3 + \cos \theta d\varphi), \\
e^{\varphi}= e^{f+h} \sin \theta d\varphi. &
\end{array}
\end{equation}
The new radial coordinate $\rho$ is related to $r$ by $d \rho = e^{-k(r)} dr$ and turns out to be convenient in subsequent computations. Written in this basis, the RR 3-form field strength becomes
\begin{align}
F_{(3)}&= -2N_c e^{-3f -2g -k} e^1 \wedge e^2 \wedge e^3+ \frac{N_c}{2} b' e^{-3f -g -h -k} e^\rho \wedge (-e^{\theta} \wedge e^1 + e^{\varphi} \wedge e^2)\nonumber\\
& \quad + \frac{N_c}{2} e^{-3f -2h -k} (a^2 -2 ab +1) e^{\theta} \wedge e^{\varphi} \wedge e^3 \\\nonumber
& \quad +N_c e^{-3f -g -h -k} (b-a) (-e^{\theta} \wedge e^2 + e^1 \wedge e^{\varphi})
\wedge e^3.
\end{align}
To be able to account for the backreaction of the flavor branes, the action has to be augmented by the DBI and Wess-Zumino actions for the flavor D5-branes. The complete action then reads
\begin{equation}
S=S_{\text{grav}}+ S_{\text{flavor}},
\end{equation}
where, in Einstein frame, we have
\begin{equation}
S_{\text{grav}}= \frac{1}{2\kappa_{(10)}^2}\int d^{10}x \sqrt{-g_{(10)}} \left( R -\frac{1}{2} (\partial_{\mu}\phi)(\partial^{\mu} \phi) - \frac{1}{12}e^{\phi}F_{(3)}^2 \right),
\end{equation}
and
\begin{equation}\label{eq:Sflavor}
S_{\text{flavor}}= T_5 \sum^{N_f} \left( - \int_{\MM_6}d^6x e^{\phi/2} \sqrt{-g_{(6)}} + \int_{\MM_6} P[C_{(6)}]\right).
\end{equation}
One of the effects of smearing the $N_f \rightarrow \infty$ flavor branes along the two transverse 2-spheres is that there will be no dependence on the angular coordinates $(\theta,\varphi,\tilde{\theta},\tilde{\varphi})$ of the functions $f(\rho), h(\rho), g(\rho)$ and $k(\rho)$ that determine our metric ansatz (\ref{eq:defmetansatz})\footnote{The validity of this will be discussed in Section \ref{sec:sols}.}, significantly simplifying the computations. After the smearing we can write
\begin{equation}
S_{\text{flavor}} =\frac{T_5 N_f}{4 \pi^2}\left( -\int d^{10}x \sin \theta \sin \tilde{\theta} e^{\phi/2} \sqrt{-g_{(6)}} + \int \text{Vol}({\mathcal Y}_4) \wedge C_{(6)}\right),
\end{equation}
with the definition $\text{Vol}({\mathcal Y}_4) = \sin \theta \sin\tilde d \theta \wedge d \varphi \wedge d \tilde{\theta} \wedge d \tilde{\varphi}$. Once the smeared flavor D5-branes are incorporated into the background, the Bianchi identity for $F_{(3)}$ (which is identical to the e.o.m. for $C_{(6)}$) gets modified to
\begin{equation}
d F_{(3)}= \frac{N_f}{4} \sin \theta \sin \tilde{\theta} d \theta \wedge d \varphi \wedge d \tilde{\theta} \wedge d \tilde{\varphi},
\end{equation}
as a result of adding a Wess-Zumino term. This can be solved by adding the following term to the original $F_{(3)}$ in (\ref{eq:F3o}),
\begin{equation}
F_{(3)}^{\text{flavor}}= -\frac{N_f}{4} \sin \theta d \theta \wedge d \varphi \wedge w^3 = -\frac{N_f}{2} e^{-3f-2h-k} e^\theta \wedge e^\varphi \wedge e^3.
\end{equation}
The full RR 3-form field strength now reads
\begin{align}
F_{(3)}&= -2N_c e^{-3f -2g -k} e^1 \wedge e^2 \wedge e^3+ \frac{N_c}{2} b' e^{-3f -g -h -k}
e^r \wedge (-e^{\theta} \wedge e^1 + e^{\varphi} \wedge e^2)\nonumber\\
& \quad + \frac{N_c}{2} e^{-3f -2h -k} (a^2 -2 ab +1-x) e^{\theta} \wedge e^{\varphi} \wedge
e^3 \\\nonumber
& \quad +N_c e^{-3f -g -h -k} (b-a) (-e^{\theta} \wedge e^2 + e^1 \wedge e^{\varphi})
\wedge e^3,
\end{align}
where the only modification is the appearance of the term involving $x=\frac{N_f}{N_c}$ in the second line.\\
The first order BPS equations for this ansatz (see Appendix B, \cite{Casero:2006pt}) can be partially solved, yielding
\begin{subequations}\label{eq:BPSsol}
\begin{align}
b(\rho) &= \frac{(2-x)\rho}{\sinh (2\rho)},\label{eq:BPSb}\\
e^{2g} &= \frac{N_c}{2}\frac{2b \cosh (2 \rho)-2+x}{a \cosh (2 \rho)-1},\\
e^{2h} &= \frac{e^{2g}}{4}(2a \cosh (2\rho) - 1 -a^2),
\end{align}
\end{subequations}
leaving us with (coupled) differential equations for $a(\rho),k(\rho)$ and $f(\rho)$
\begin{subequations}\label{eq:BPS}
\begin{align}
\partial_{\rho} a &= \frac{2}{2\rho \coth 2\rho}\left( -
\frac{2e^{2 k}-N_f}{2N_c-N_f}\frac{(a \cosh 2\rho -1)^2}{\sinh 2\rho}+a^2 \sinh 2\rho -2a \rho\right),\\
\partial_{\rho} k &=\frac{2}{(2\rho \coth 2\rho)(1-2a \cosh 2\rho + a^2)}
\left(\frac{2e^{2 k}+N_f}{2N_c-N_f}a \sinh 2\rho (a \cosh 2 \rho -1) + \right.\nonumber\\
&\left. \quad + 2\rho (a^2\frac{\sinh 2\rho}{2\rho}\cosh 2\rho -2a \cosh 2\rho +1)\right) ,\\
\partial_{\rho} f &= \frac{(-1+a \cosh 2\rho)\sinh^{-2}2\rho}{4(1+a^2-2a \cosh 2\rho)(-1+2\rho \cosh 2\rho)}
 \nonumber\\
& \quad \times \left(-4\rho +\sinh 4\rho + 4a\rho \cosh 2\rho -2a \sinh 2\rho -
\frac{4}{2-x}a\sinh^3 2\rho \right)\,.
\end{align}
\end{subequations}
Equations (\ref{eq:BPS}) can be solved numerically for $x \neq 2 $. A detailed discussion of the solutions along with asymptotic expansions can be found in~\cite{Casero:2006pt},\cite{Casero:2007jj}. Note that the equations~(\ref{eq:BPS}) are not valid for $x=2$. In this case (cf. Appendix D of \cite{Casero:2006pt}) equation \eqref{eq:BPSb} implies that in the limit $x \rightarrow 2$ we have $b(\rho)=0$. Furthermore, the authors in \cite{Casero:2006pt} observe from their numerical solutions that $a(\rho)$ asymptotes to  $1/\cosh{2 \rho}$ in the limit $x \rightarrow 2$. In the Appendix \ref{ap:BPS-eqs} we derive the BPS equations for $b(\rho)=0$. For 
\begin{equation}
a=\frac{1}{\cosh 2\rho},
\end{equation}
they reduce to
\begin{subequations}\label{eq:diffeq}
\begin{align}
e^{h-g} &= \frac12 \tanh{2 \rho}\,,\\
\partial_{\rho} e^{h+g} &= e^{2k} -N_c\,,\\
\partial_{\rho} k &= -(e^{2k}+N_c)e^{-h-g}+2 \coth 2\rho\,,,\\
\partial_\rho \phi &= 4 \partial_{\rho} f= N_c e^{-g-h}\,.
\end{align}
\end{subequations}
Solving this system of equations will be the topic of Subsection \ref{subsec:sols} below.\\
As an aside, let us also briefly discuss yet another solution to the BPS equations (\ref{eq:BPSsol}) and (\ref{eq:BPS}) for $x=2$ that was found in \cite{Casero:2006pt} by embedding a large number of flavor branes in the singular CVMN background and will be important later. As discussed above, from equation \eqref{eq:BPSb} it follows that $b(\rho)=0$ and it is consistent to set $a(\rho)=0$ as well.\footnote{The two classes of solutions could be named abelian for $a(\rho)=0$ and non-abelian for $a(\rho)=1/\cosh 2\rho$ in analogy with the abelian and non-abelian CVMN backgrounds. We expect that there should again be a one-parameter family of solutions interpolating between the two, analogous to the one-parameter family interpolating between the abelian and non-abelian CVMN background found in \cite{Gubser:2001eg}.}   In \cite{Casero:2006pt} the solution to the BPS equations is found to be
\begin{equation}\label{eq:xi2}
e^{2h}=\frac{N_c}{\xi},\quad e^{2g}=\frac{N_c}{4-\xi},\quad e^{2k}=N_c,\quad \phi = 4f=\phi_0 + \rho,
\end{equation}
where $0 < \xi <4$. The corresponding Einstein frame metric has a singularity for $\rho \rightarrow 0$. However, this singularity can be smoothed out by turning on a temperature in the dual field theory, thus finding a black hole solution. The black hole temperature turns out \cite{Bertoldi:2007sf} to be reminiscent of the Hagedorn temperature in little string theory. Therefore, it seems that this solution is not dual to a field theory but to a little string theory.\\

\section{New Supergravity Solutions}\label{sec:sols}

In this section we consider the inclusion of a large number of flavor branes into the CVMN background for the special case of $N_f=2N_c$. We discuss the validity and shortcomings of the smearing approach in Subsection~\ref{sub:incl} before solving the BPS equations \eqref{eq:diffeq} in Subsection~\ref{subsec:sols}. In Subsection~\ref{decouplinglimit} we discuss properties of our solutions.

\subsection{Including a Large Number of Flavor Branes}\label{sub:incl}

We take the flavor branes to extend along the $x^\mu, \psi, \rho$ directions and place them at fixed values in the $\theta, \varphi, \tilde \theta, \tilde \varphi$ directions. The reason for taking the branes to extend along the non-compact $\rho$ direction is that we want to send their 't Hooft coupling to zero. For our flavor branes we have
\begin{equation}
\lambda_f=\frac{2 (2 \pi)^2 \sqrt{\alpha' g_s N_c}}{Vol(S^1)},
\end{equation}
where $Vol(S^1)$ is the dimensionless volume of the $\psi$-direction wrapped by the branes. For energies small compared to the inverse size of the $S^1$ this coupling determines the interaction of the 5-dimensional gauge fields living on the flavor branes and the quarks living on the 4-dimensional intersection of the color and flavor branes.\footnote{It should be noted that the symmetry is formally a global symmetry due to the presence of a non-compact direction but there still is a non-vanishing interaction between the quarks due to their charge under this symmetry unless $\lambda_f=0$.} When taking the near horizon limit we take $\alpha' \rightarrow 0$ to decouple open and closed string modes while keeping $g_s N_c$ large but finite to be able to trust the supergravity approximation. In this limit the interaction between the quarks and the five dimensional gauge fields disappears and we obtain SQCD with a global flavor symmetry. Note that this would not be possible if the flavor branes wrapped a compact 2-cycle.\\
As mentioned above, we have to include the DBI and Wess-Zumino actions \eqref{eq:Sflavor} for the flavor branes. Using delta functions these can be written as integrals over the 10-dimensional spacetime. However, solving the equations of motions containing delta functions is hard. So we invoke a 'smearing' process first proposed in \cite{Bigazzi:2005md}. If this smearing process is to preserve supersymmetry, we have to ensure that all the branes at different positions in the transverse space preserve the same supersymmetry as the background. This was checked in~\cite{Nunez:2003cf} and can easily be seen from the results in Appendix \ref{ap:BPS-eqs}. Formally, this smearing process can be thought of as follows: As is well known the delta function can be written as an infinite sum using the completeness relation for the eigenfunctions of the Laplacian on the space transverse to the branes. The smearing corresponds to truncating this sum to the zero mode. Intuitively, however, one might want to evenly distribute a very large number $N_f$ of branes over their transverse directions, {\it i.e.}, in our case the $\theta, \varphi, \tilde \theta, \tilde \varphi$ directions. It is not immediately apparent that these two prescriptions are equivalent. To check this, one should make an ansatz for all the fields in the type IIB action, the DBI action, and the Wess-Zumino action that depends on the positions of all the branes. Then one can expand all fields and the delta functions in two sets of spherical harmonic functions corresponding to the two sets of coordinates $\theta, \varphi$ and $\tilde \theta, \tilde \varphi$. The hope then is that there exists a consistent truncation to the zero modes and that in the large $N_f$ limit the contribution of the higher modes is negligible. This seems plausible due to the homogeneous distribution of a very large number of flavor branes. In this paper we do not attempt to verify this hope but rather try to provide some checks that indicate that the interpretation of our solutions as being dual to $\mathcal N=1$ SQCD might be consistent. Note that our solutions are supergravity solutions with $\mathcal{N}=1$ supersymmetry since we solve the BPS equations. However, to conjecture that they are dual to a gauge theory would require that the smearing prescription provides a valid approximation to the brane configuration.\\
Before we move on, some more comments on the consistency of our setup are in order. The flavor branes extend along all non-compact directions. Gauss's law then implies that these branes cannot carry a net charge. As in \cite{Karch:2002sh}, we accomplish this in our setup by wrapping the branes around a topologically trivial cycle in the compact space that carries no flux. To be more specific, our flavor branes wrap the Hopf fiber, {\it i.e.} an $S^1$ inside the $S^3$, which is clearly trivial. Most stable D-brane configurations are stable precisely due to the charges they carry. So we should ask if these flavor branes will not simply slip off the $S^3$. A stability analysis shows that the slipping mode for a probe brane embedded in this way into the CVMN background indeed has a tachyonic potential as expected but is in fact stabilized by effects similar to the ones observed in \cite{Karch:2000gx},\cite{Bachas:2000fr}. In other words there exists the analogue of a Breitenlohner-Freedman bound in this geometry. Furthermore, one can show \cite{Nunez:2003cf} that the embedding is kappa symmetric, which implies that the configuration we would like to smear is not only stable but also supersymmetric \cite{Bergshoeff:1997kr}.\footnote{The fact that the brane probe is a solution of the worldvolume equations of motion together with kappa symmetry already implies supersymmetry.} For the interested reader the stability analysis is carried out in Appendix \ref{ap:stability}.\\
To verify that the new background in which the backreaction of the smeared flavor branes is taken into account still preserves the same amount of supersymmetry as the CVMN background, we show in Appendices \ref{ap:stability} and \ref{ap:BPS-eqs} that an additional probe brane extending along $x^\mu, \psi, \rho$ and at fixed $\theta, \varphi, \tilde \theta, \tilde \varphi$ is still both stable under small fluctuations and kappa symmetric implying that the system is indeed supersymmetric \cite{Bergshoeff:1997kr}.\\
Since we are placing our flavor branes at different positions in the transverse space they support a $U(1)^{N_f}$ and not a $U(N_f)$ gauge group. However, at energies small compared to the inverse size of the compact transverse space the theory will not be able to resolve the transverse space and should effectively behave as a theory with $U(N_f)$ symmetry. So our compact directions need to be big enough to trust supergravity but also small enough so that, at the energies we are interested in, we can neglect KK modes and have a $U(N_f)$ flavor symmetry.

\subsection{Solutions of the BPS Equations for $N_f=2N_c$}\label{subsec:sols}

\begin{figure}[h]
\begin{center}
\includegraphics[width=5in]{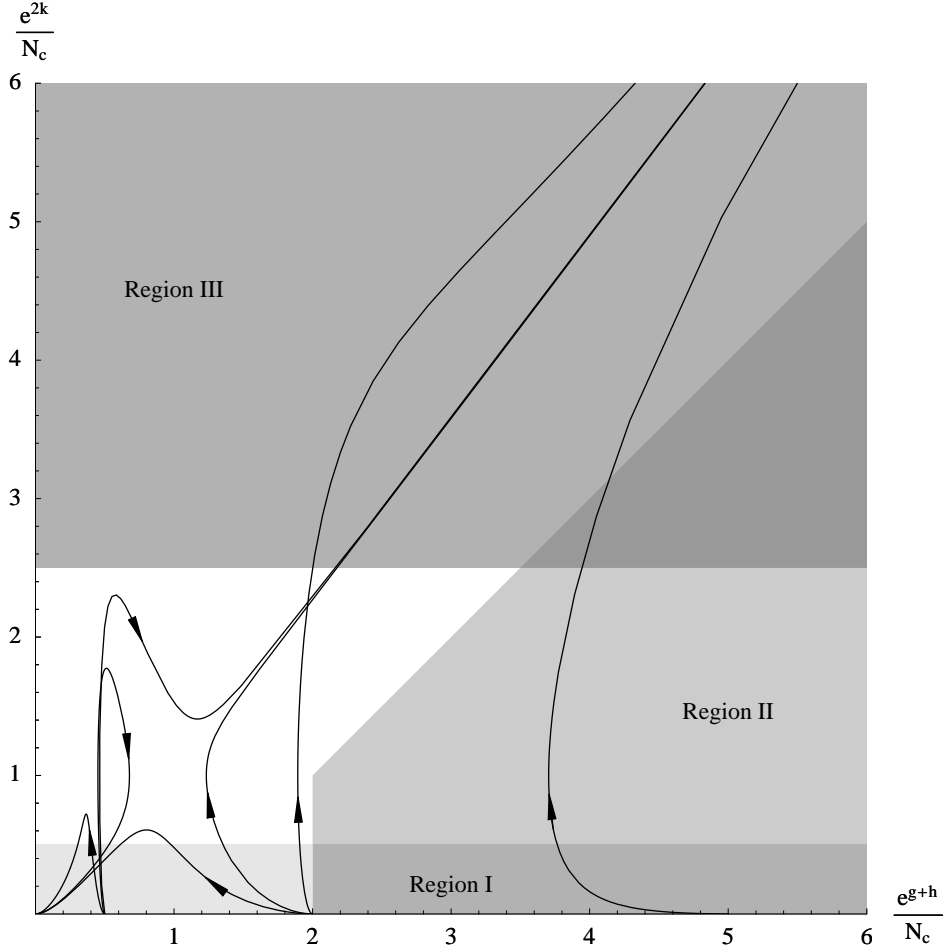}
\end{center}
\caption{This plot shows examples of numerical solutions of the above system of equations \protect\eqref{eq:diffeq2} as well as the regions in which the equations can be solved analytically. We plot three solutions starting at both $e^{g+h}/N_c= \rho_0=1/2$ and $e^{g+h}/N_c=\rho_0=2$ and one starting at $e^{g+h}/N_c=\rho_0=5$.} \label{fig:regions}
\end{figure}

\begin{figure}[h]
\begin{center}
\includegraphics[width=5in]{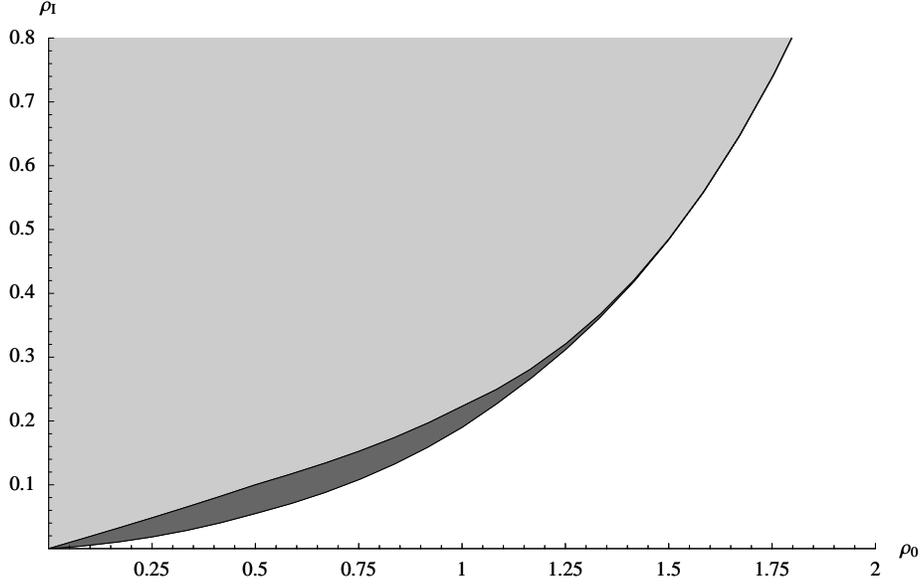}
\end{center}
\caption{This plot shows the regions in the $\rho_0-\rho_I$ parameter space corresponding to different classes of solutions. Both shaded regions correspond to the solutions with a singularity at finite $\rho$. For the region shaded in lighter gray $e^{g+h}$ is monotonically decreasing. The region shaded in darker gray corresponds to the solutions describing a loop in the $e^{2k}$-$e^{g+h}$ plane. The white region corresponds to solutions that approach $e^{2 k}=\frac{4}{3} e^{g+h} -\frac{N_c}{2}$ for large $\rho$.} \label{fig:rho0rhoI}
\end{figure}

In this subsection we present new supergravity solutions describing $N_f= 2 N_c$ smeared flavor branes in the CVMN background, extending the result in Appendix D of \cite{Casero:2006pt}.\\
Our metric ansatz is \eqref{eq:defmetansatz}
\begin{align}\label{eq:metric}
ds_{10}^2 &= e^{2f(\rho)} \Big(dx_{1,3}^2+ e^{2k(\rho)} d\rho^2
+ e^{2h(\rho)} (d\theta^2 + \sin^2 \theta d\varphi^2)\nonumber\\
& \quad +\frac{e^{2g(\rho)}}{4} \left( (w^1 +a(\rho)d\theta)^2 +(w^2 -a(\rho) \sin \theta d \varphi)^2
\right) + \frac{e^{2k(\rho)}}{4}(w^3 + \cos \theta d \varphi)^2 \Big),
\end{align}
where $a(\rho)=\frac{1}{\cosh{2\rho}}$ and the $w^a$ are given in \eqref{eq:leftinv}. The RR 3-form in the vielbein basis \eqref{eq:vielbein} is
\begin{align}\label{eq:3form}
F_{(3)}&= -2N_c e^{-3f -2g -k} e^1 \wedge e^2 \wedge e^3 + \frac{N_c}{2} e^{-3f -2h -k} (a^2 -1) e^{\theta} \wedge e^{\varphi} \wedge e^3 \nonumber\\
& \quad +N_c\, a \,e^{-3f -g -h -k} (e^{\theta} \wedge e^2 + e^{\varphi} \wedge e^1) \wedge e^3.
\end{align}
The BPS equations are derived in Appendix \ref{ap:BPS-eqs} and reduce to $e^{h-g}=\frac12 \tanh{2\rho}$,
\begin{subequations}\label{eq:diffeq2}
\begin{align}
\partial_{\rho} e^{h+g} &= e^{2k} -N_c,\label{eq:diffehg}\\
\partial_{\rho} k &= -(e^{2k}+N_c)e^{-h-g}+2 \coth 2\rho,\label{eq:diffek}
\end{align}
\end{subequations}
and
\begin{equation}\label{eq:dilaton}
\partial_\rho \phi = 4 \partial_{\rho} f= N_c e^{-g-h}.
\end{equation}
Note that the $N_c$ dependence of the solution is trivial since rescaling $e^{2k}$ and $e^{g+h}$ by a factor of $N_c$ eliminates $N_c$ from the equations.\\
Once a solution to the coupled first order equations \eqref{eq:diffeq2} is found the dilaton $\phi$ can be obtained by integration of \eqref{eq:dilaton}. The equations \eqref{eq:diffeq2} do not seem to allow us to write the {\it general} solution in a closed analytic form. The equations can easily be solved numerically and there are in fact a number of things that can be done analytically. We can find an analytic solution valid for small values of $\rho$ in the form of a power series. To leading order the solution takes the form
\begin{eqnarray}
e^{g+h}/N_c&=&\rho_0-\rho,\\
e^{2k}/N_c&=&\left(\frac{\rho}{\rho_I}\right)^2,
\end{eqnarray}
implying that, at least for all solutions of this form, $e^{2k}$ goes to zero for small $\rho$.\\
Instead of studying the small $\rho$ behavior we can find closed analytic expressions approximating the two-parameter family of solutions to the equations \eqref{eq:diffeq2} in the following regions:
\begin{align}\label{eq:apregions}
&\text{Region I:}    &e^{2k}&\ll N_c\nonumber\\
&\text{Region II:}   &e^{g+h}&\gg N_c, e^{2k}\\
&\text{Region III:}  &e^{2k}&\gg N_c\nonumber
\end{align}
These regions, together with some examples of numerical solutions, can be seen in Figure\nolinebreak~\ref{fig:regions}.
There are two qualitatively different two-parameter families of solutions. For small values of $\rho$ both start out at $e^{2k}=0,\, e^{g+h}/N_c=\rho_0$. For large $\rho$ one of the families crosses $e^{2 k}=\frac{4}{3} e^{g+h} -\frac{N_c}{2}$ and then approaches it from above. The other family reaches a singularity at $e^{2k}=e^{g+h}=0$ for a finite value of $\rho$.
For $\rho_0 < \rho_{c} \sim 1.6$, we can make a further distinction among the singular solutions. For given $\rho_0< \rho_c$ and large enough values of $\rho_I$ the function $e^{g+h}$ decreases monotonically while below a certain value of $\rho_I$ the solutions describe a loop in the $e^{2k}$-$e^{g+h}$ plane. A schematic plot showing the different regions in the $\rho_0$-$\rho_I$ parameter space corresponding to the three classes of solutions is shown in Figure \ref{fig:rho0rhoI}.
The singular and non-singular two-parameter families are separated by a one-parameter family that approaches $e^{2k} = e^{g+h} = N_c$ for $\rho \rightarrow \infty$. In fact, it asymptotically approaches the $\xi=2$ solution found in \cite{Casero:2006pt} and given above in equation \eqref{eq:xi2}. This one-parameter family will play a special role as discussed in the next subsection.\\
The analytic solutions discussed below comprise two-parameter families originating from Region I as long as they remain inside Regions I, II or III for all $\rho$ as well as a one-parameter family contained entirely in Region III. We will merely summarize these analytic solutions in this section and refer the interested reader to Appendix \ref{ap:solutions} for a detailed derivation.\\
Unfortunately, we were not able to find an analytic solution for the one-parameter family separating the two two-parameter families. We will therefore have to study them numerically in the next subsection. \footnote{One particular solution of the one-parameter family of solutions ending at the point $e^{2k} =e^{g+h} =N_c$ and coming from the overlap of Region I and II was already found in Appendix D of \cite{Casero:2006pt}.} \\

{\bf The Solution Contained in Region I}\\[.3cm]
For $e^{2k} \ll N_c$ we can solve the BPS equations \eqref{eq:diffeq2} to first order in $e^{2k}$. The solution is
\begin{eqnarray}
e^{g+h}/N_c&=&\rho_0-\rho+\frac{1}{64{\rho_0}^2{\rho_I}^2}\left(\frac{1}{4}
\left(1+8(\rho-\rho_0)^2\sinh{4\rho}\right)\right.\nonumber\\
&&-\left.(\rho-\rho_0)\cosh{4\rho}-\rho_0+\frac{8}{3}\rho(\rho^2-3\rho\rho_0 +3\rho_0^2)\right),\\
e^{2k}/N_c&=&\left(\frac{\sinh{2\rho}}{2\rho_I}\left(1-\frac{\rho}{\rho_0} \right)\right)^2.
\end{eqnarray}
For this solution to be entirely contained in Region I we need to impose $e^{2k} \ll N_c$ for all $\rho$. Note that this solution does not only have a singularity at $\rho=0$ but also a second one near $\rho=\rho_0$. At the singularity at $\rho=0$ the dilaton and the metric component $g_{tt}$ both go to a constant. On the other hand both go to infinity at the other singularity. It is not clear whether these solutions can be given a physical interpretation.\\

{\bf The Solution Starting in Region II and Ending in Region III}\\[.3cm]
For a somewhat limited choice of initial conditions, meaning $\rho_0 \gg 1$ and $\rho_I \ll 1$ or $\rho_0, \rho_I \gg 1$, we can find a closed form for the analytic solution approximating the solution starting in the overlap of Region I and II and ending in Region III.
For $\rho_I \ll 1$ and $\rho_0 \gg 1$ we find
\begin{eqnarray}
e^{g+h}/N_c&=&\rho_0\left(1-\frac{1}{8\rho_0{\rho_I}^2}\left(3\rho-\frac{3}{4}\sinh{4\rho}\right)\right)^{\frac{1}{3}},\\
e^{2k}/N_c&=&\left(\frac{\sinh{2\rho}}{2\rho_I}\right)^2\left(1-\frac{1}{8\rho_0{\rho_I}^2}\left(3\rho- \frac{3}{4}\sinh{4\rho}\right)\right)^{-\frac{2}{3}},
\end{eqnarray}
while for $\frac14 e^{2 \rho_0} \gg \rho_I \gg 1$ and $\rho_0 \gg 1$ the solution is
\begin{eqnarray}
e^{g+h}/N_c&=&\rho_0\left(1+\frac{3e^{4\rho}}{64\rho_0{\rho_I}^2}\right)^{\frac{1}{3}}-\rho,\\
e^{2k}/N_c&=&\left(\frac{\sinh{2\rho}}{2\rho_I}\right)^2\left(1+ \frac{3e^{4\rho}} {64\rho_0{\rho_I}^2} \right)^{-\frac{2}{3}}.
\end{eqnarray}
As discussed below, we think that these solutions describe the geometry generated by the color and flavor D5-branes.
For somewhat less restrictive initial conditions the matching can still be done but with piecewise defined functions that give a slightly better approximation to the numerical results. This is described in more detail in Appendix \ref{ap:solutions}.\\

{\bf The Solution Contained in Region III}\\[.3cm]
Solving the equations for $e^{2k} \gg N_c$ by neglecting $N_c$ in the BPS equations \eqref{eq:diffeq2} one finds
\begin{eqnarray}
e^{g+h}/N_c&=&A\left(8B-12\rho+3\sinh{4\rho}\right)^{\frac{1}{3}},\\
e^{2k}/N_c&=&\frac{8A\sinh^2{2\rho}}{(8B-12\rho+3\sinh{4\rho})^{\frac{2}{3}}}.
\end{eqnarray}
For this solution to be entirely contained in Region III we need to demand that $e^{2k} \gg N_c$ for all $\rho$. In particular this has to hold in the limit $\rho \rightarrow 0$ which implies $B=0$. After making the identification $A=\frac{\epsilon^\frac{4}{3}}{2^\frac{4}{3}3^\frac{1}{3}}$ this solution can be seen to correspond to the deformed conifold\nolinebreak~\cite{Candelas:1989js}
\begin{eqnarray}
e^{g+h}/N_c&=&\frac{\epsilon^\frac{4}{3}\left(\sinh{4\rho}-4\rho\right)^{\frac{1}{3}}}{2^\frac{4}{3}},\\
e^{2k}/N_c&=&\frac{2}{3}\epsilon^\frac{4}{3}\frac{2^\frac{2}{3}\sinh^2{2\rho}}{(\sinh{4\rho}-4\rho)^{\frac{2}{3}}}.
\end{eqnarray}
In order for it to be contained entirely in Region III we require $\epsilon\gg \left(\frac{3}{2}\right)^\frac{1}{4} \approx 1$.

\subsection{Discussions of the Solutions and the Decoupling Limit}\label{decouplinglimit}

\begin{figure}[!h]
\begin{center}
\includegraphics[width=5.5in]{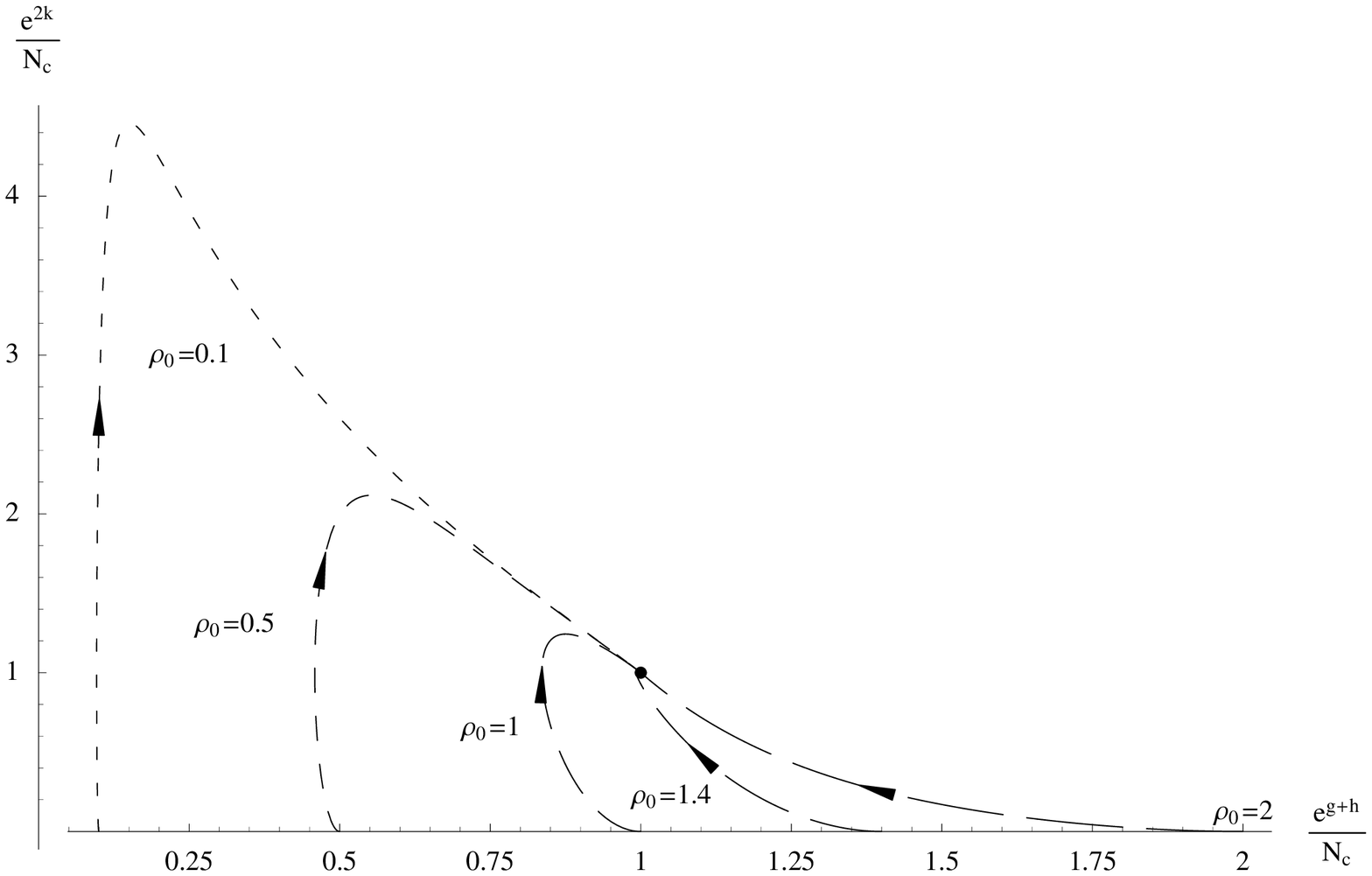}
\includegraphics[width=6.3in]{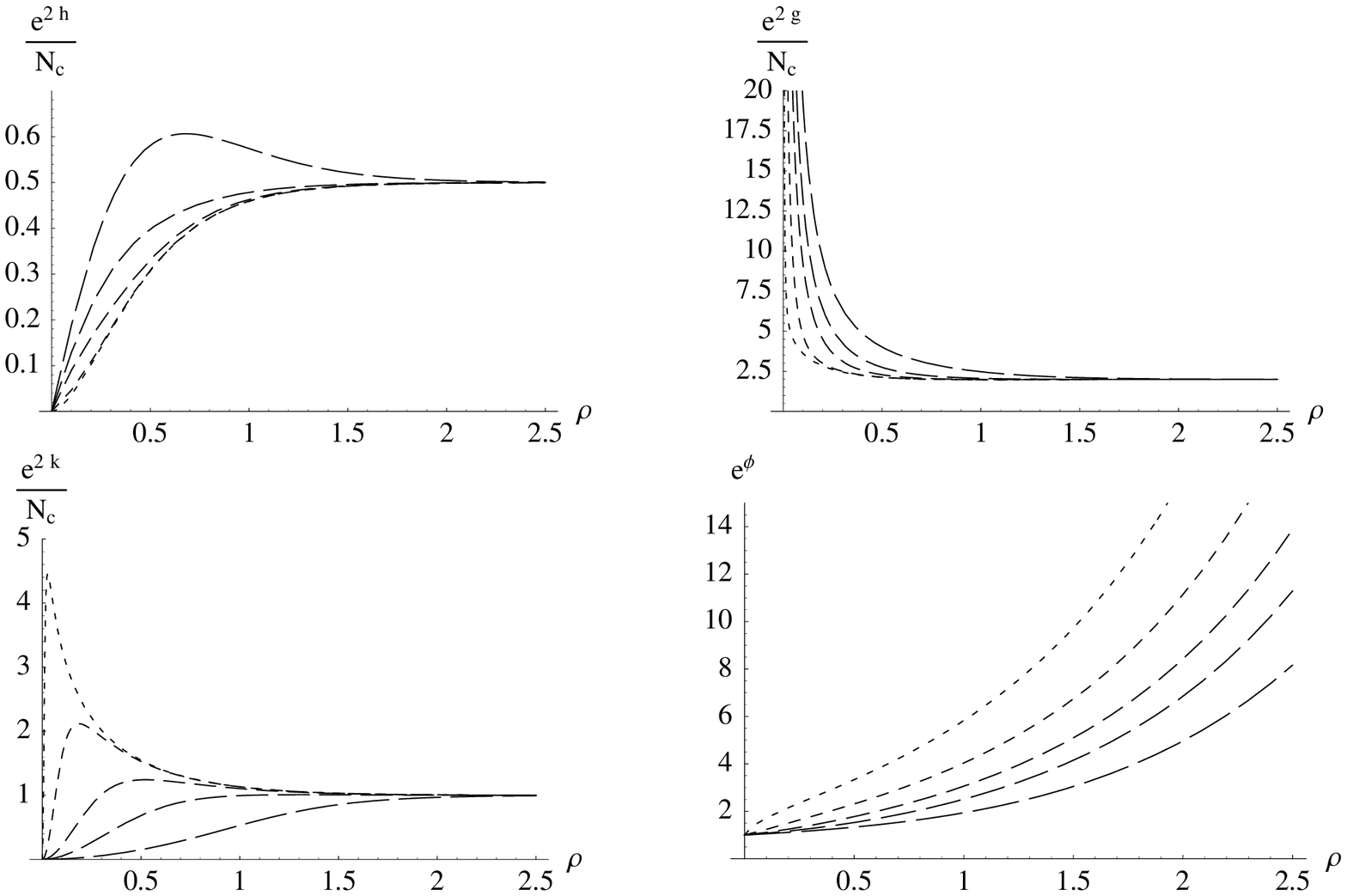}
\end{center}
\caption{This plot shows representatives of our one-parameter family of solutions separating the singular and non-singular two-parameter families starting at  $e^{g+h}/N_c=\rho_0=0.1,0.5,1,1.4,2$ and $e^{2k}=0$. Longer dashes correspond to larger $\rho_0$ values.} \label{fig:one-parameter}
\end{figure}

The two-parameter families of solutions reaching the point $e^{2k}=e^{g+h}=0$ for finite $\rho = \rho_s$ are singular both at $\rho=0$ as well as at $\rho=\rho_s$. The Ricci scalar close to the singularities scales as $1/\rho^2$ and $1/(\rho - \rho_s)^2$, respectively. We do not know whether these solutions are physical.\\
For the two-parameter family of solutions which approaches $e^{2 k}=\frac{4}{3} e^{g+h} -\frac{N_c}{2}$ as $\rho \rightarrow \infty$ we can write the metric for large $\rho$ as
\begin{equation}
ds^2 = \alpha' g_s N_c \, e^{\phi(r)/2} \left( \frac{dx_{1,3}^2}{\alpha' g_s N_c} +dr^2 + r^2 ds^2_{T_{11}} \right),
\end{equation}
where we defined a new radial coordinate $r \sim e^{2\rho/3}$ and $ds^2_{T_{11}}$ is the Einstein metric on the homogeneous space $(SU(2) \times SU(2))/U(1)_{\text{diag}}$. So for large values of the radial coordinate the metric approaches a product of Minkowski space and the singular conifold. As $\rho$ goes to zero the geometry is singular and the Ricci scalar diverges like $1/\rho^2$. This should be compared to the usual $1/\rho$ divergence for flat D5-branes. We attribute the discrepancy to the way the flavor branes are embedded. According to the criteria developed in \cite{Maldacena:2000mw} the solutions should still be accepted as being physical since the $g_{tt}$ component of the metric remains bounded as $\rho$ approaches zero. The dilaton increases monotonically with $\rho$ and for large values of $\rho$ approaches a constant. So as long as $g_s N_c \gg 1$ the Ricci scalar is small for all large enough $\rho$, and, if $g_s \ll 1$ as well, this means that the supergravity approximation is valid. We believe that these solutions describe the geometry generated by the $N_c$ color branes in the presence of $N_f = 2 N_c$ smeared flavor branes.\\
These two two-parameter families are separated by a one-parameter family that approaches $e^{2k} = e^{g+h} = N_c$ for $\rho \rightarrow \infty$. This situation of having two classes of solutions separated by `special' solutions should be familiar from e.g. the Klebanov-Tseytlin solution \cite{Klebanov:2000nc}. There one has an integration constant $b_0$. For $b_0<0$ one finds solutions that have a second singularity for finite value of the radial coordinate. For $b_0 >0$ the solutions describe the geometries generated by the brane configuration. For $b_0=0$ one obtains special solutions describing the corresponding near horizon geometries. Before fixing the boundary conditions at infinity a similar behavior can be observed for a stack of Dp-branes in flat space. This leads us to the conclusion that our one-parameter family separating the singular and non-singular two-parameter families describes the near horizon geometry for our brane setup and should be dual to $\mathcal{N}=1$ SQCD with $N_f = 2 N_c$. Due to the singularity at $\rho=0$ we should impose an IR cut-off. There also has to be a UV cut-off due to the presence of Kaluza-Klein modes. These become important as the energy increases. At even higher energies the theory becomes six dimensional and has a little string theory as its UV completion. We should note that it is not clear that the IR and UV cut-off are even compatible (in the sense that all approximations are valid) but we find that at least for a certain range of initial conditions this is the case. In the next section we study Wilson loops and beta functions for these solutions to support our interpretation. Unfortunately, we do not have an analytic expression describing these geometries because the near horizon limit takes us out of the regime where our approximations \eqref{eq:apregions} are valid. So we study these solutions numerically. Figure \ref{fig:one-parameter} shows a few examples of this one-parameter family of solutions.

\section{Field Theory}\label{sec:sqcd}

In this section, we will use the gauge/gravity duality to study rectangular Wilson loops as well as beta functions of the field theories dual to our supergravity solutions.

\subsection{Wilson Loops}
\begin{figure}[!h]
\begin{center}
\includegraphics[width=4.3in]{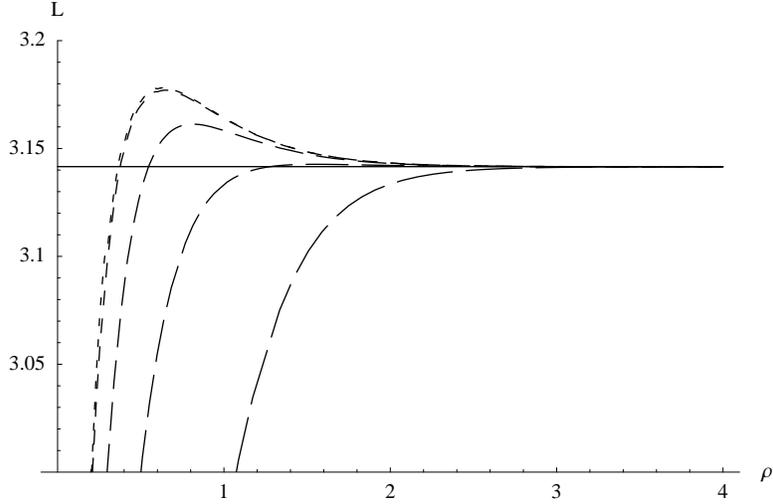}
\caption{This plot shows the separation between the endpoints of the string $L$ as a function of the minimal distance between the string worldsheet and the singularity at $\rho=0$. $L$ has units $\sqrt{\alpha' g_s N_c}$. We have dashed the lines so that longer dashes correspond to larger $\rho_0$ values and $\rho_0$ takes the values $\rho_0=0.1,0.5,1,1.4,2$. We see that $L$ has a maximum for all $\rho_0 < \rho_c \sim 1.6$.}\label{fig:lvsr}
\end{center}
\end{figure}

\begin{figure}[!h]
\begin{center}
\includegraphics[width=4.5in]{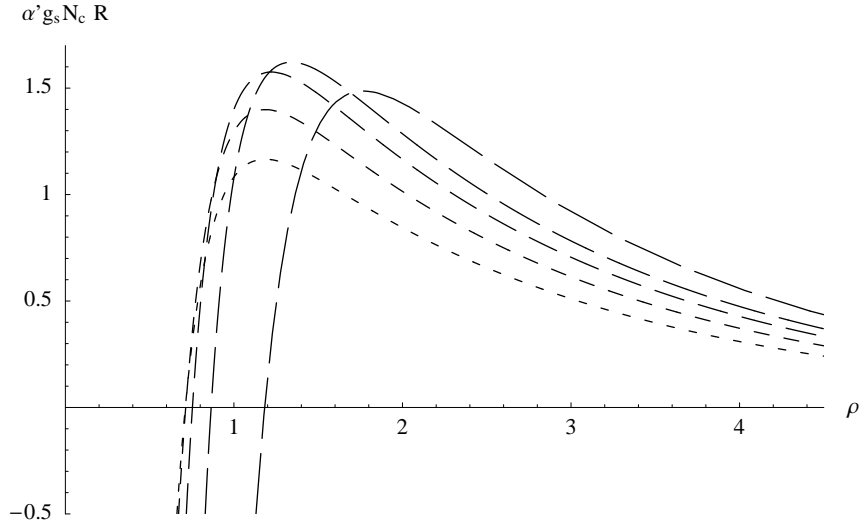}
\caption{This plot shows $\alpha' g_s N_c \, R$ versus $\rho$ for the near horizon geometries for $\rho_0=0.1,0.5,1,1.4,2$. Again longer dashes correspond to larger $\rho_0$ values. The Ricci scalar has a maximum at a $\rho$ value which is comparable to the value for which $L$ has its maximum if $\rho_0 < \rho_c \sim 1.6$.} \label{fig:Ricci}
\end{center}
\end{figure}

\begin{figure}[!h]
\begin{center}
\includegraphics[width=5.5in]{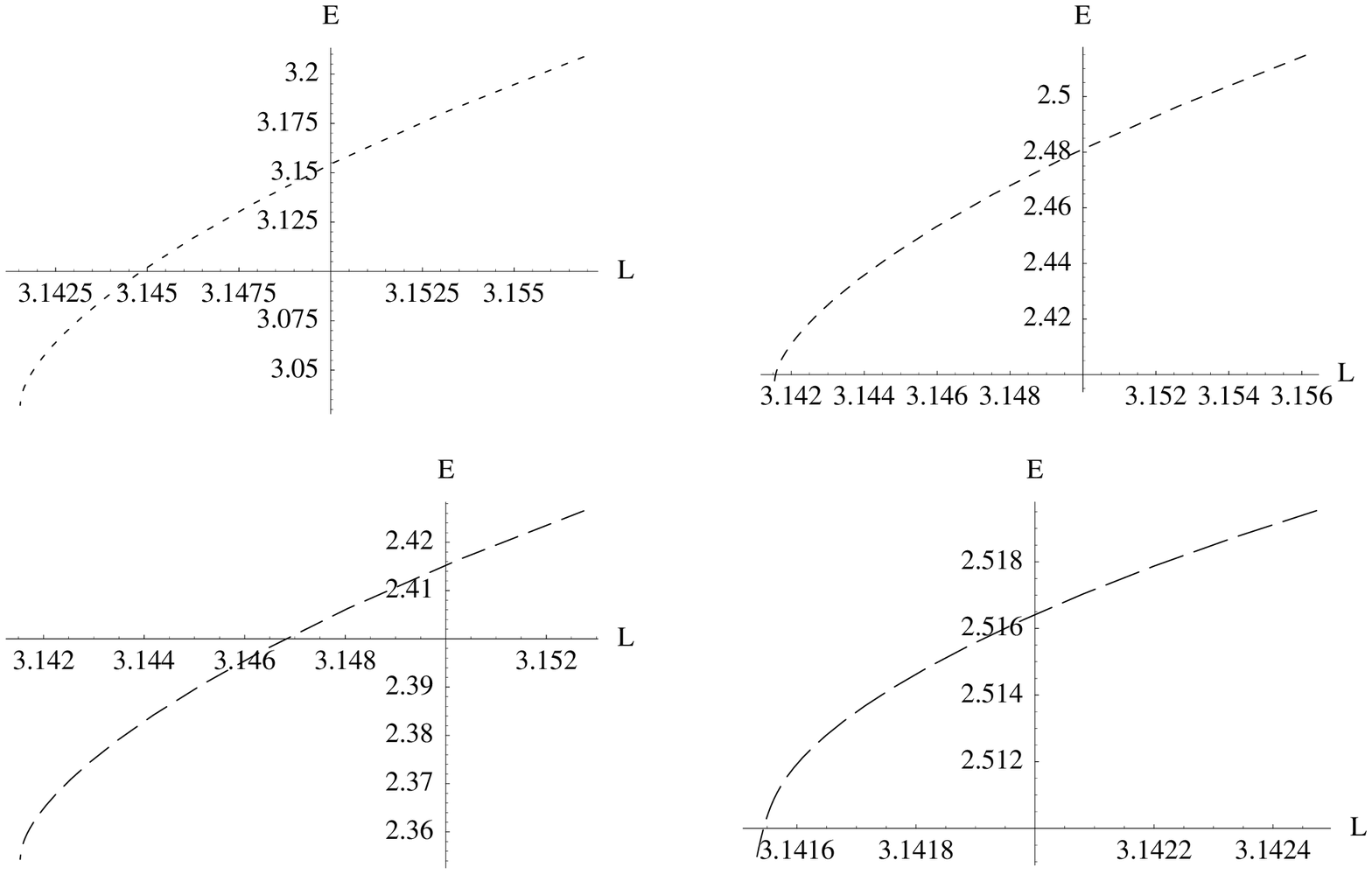}
\caption{The quark-antiquark potential for $\rho_0=0.1, 0.5,1, 1.4$. $E$ is given in units of $\frac{e^{\phi_0} \sqrt{g_s N_c}}{2 \pi \sqrt{\alpha'}}$ and $L$ in units of $\sqrt{\alpha' g_s N_c}$.}\label{fig:wilsonevsl}
\end{center}
\end{figure}

As usual the Wilson loop for two non-dynamical quarks can be computed by evaluating the Nambu-Goto action of a fundamental string with appropriate boundary conditions on shell~\cite{Maldacena:1998im},\cite{Rey:1998ik}. We will parametrize the string world sheet by $t=\tau$, $x=\sigma$ and $\rho=\rho(\sigma)$. Suppose we study a quark and an antiquark separated by a distance $L$ along the $x$-direction in gauge theory coordinates. The energy and length of the string can be given as functions of $\rho$, where $\rho$ is the minimal coordinate distance between the worldsheet and the singularity. In the string frame these become
\begin{align}\label{eq:LE}
L(\rho) &= 2 \sqrt{\alpha' g_s} \int_{\rho}^{\overline{\rho}}e^{k(\rho')}
\frac{e^{\phi(\rho)}}{\sqrt{e^{2\phi(\rho')}-e^{2\phi(\rho)}}} d\rho' ,\\
E(\rho) &= \frac{\sqrt{g_s}}{2\pi \sqrt{\alpha'}}\left(2 \int_{\rho}^{\overline{\rho}}
\frac{e^{2\phi(\rho')+k(\rho')}}{\sqrt{e^{2\phi(\rho')}-e^{2\phi(\rho)}}} d\rho'
- 2 \int_0^{\overline{\rho}} e^{\phi(\rho')+k(\rho')}d\rho' \right),
\end{align}
where $\overline{\rho}$ is a cutoff that can be taken to infinity at the end of the calculation. We will study these integrals numerically and must work with a large but finite cut-off. We have checked that the cut-off is large enough to make our results cut-off independent. The results can be used to obtain a relation $E(L)\equiv V_{q\bar{q}}(L)$. We will calculate the Wilson loops for the same set of parameters as shown in  Figure \ref{fig:one-parameter} above. For solutions with $\rho_0 < \rho_c \sim 1.6$ we find that $L(\rho)$ has a maximum at $\rho$ comparable to the value of $\rho$ for which the supergravity approximation breaks down. This is shown in Figures \ref{fig:lvsr} and \ref{fig:Ricci}. For $\rho_0 > \rho_c \sim 1.6$ $L(\rho)$ approaches $\sqrt{\alpha' g_s N_c} \pi$ from below.\footnote{Using the large $\rho$ behavior of our solutions, namely that $e^{2k} \rightarrow N_c$ and $e^\phi \rightarrow e^{\phi_0 + \rho}$, it is straightforward to check that for large $\rho$ we have $L(\rho) \sim 2 \sqrt{\alpha' g_s N_c} \int_0^\infty \frac{d\rho'}{\sqrt{e^{2\rho'}-1}} = \pi \sqrt{\alpha' g_s N_c}$ .}
We can only trust the results of our calculations in a regime where the string worldsheet is entirely contained in regions where the supergravity approximation is valid and the KK-modes are negligible. The scale at which the KK-modes become important was identified in \cite{Casero:2006pt} as the value of $\rho$ for which $L(\rho)$ reaches the value $L(\infty)= \pi \sqrt{\alpha' g_s N_c}$. For all solutions this occurs around $\rho \simeq 2.5$. The value of $\rho$ above which the curvature is small and supergravity can be trusted depends on the particular solution. The Ricci scalar for some representatives can be seen in Figure \ref{fig:Ricci}. For $\rho_0 < \rho_c$ the curvature is small for $\rho \gtrsim 1.2$ while for large $\rho_0$ we can only trust supergravity for $\rho \gtrsim \rho_0$. Note that this means that for $\rho_0 \gtrsim \rho_c \sim 1.6$ there is no regime in which both supergravity can be trusted and the KK-modes can be neglected. For small $\rho_0$ we can trust our results for $1.2 \lesssim \rho \lesssim 2.5$. The quark-antiquark potential extracted from these Wilson loops for this range of the radial variable is shown in Figure \ref{fig:wilsonevsl}.
Over the range of radial variable $\rho$ for which our approximations are valid the function $L(\rho)$ is a monotonic function of $\rho$ and we do not observe any signs of pair creation. We should not try to interpret the maximum in $L(\rho)$ in any way since it occurs for values of $\rho$ where the supergravity approximation is no longer valid and one should expect stringy corrections to alter the behavior of $L(\rho)$ for small $\rho$. We do think that the dual field theory contains dynamical quarks and their effects should certainly be observable on the supergravity side but we think that pair creation in our models occurs in a regime where stringy corrections can no longer be neglected. One might speculate that $L(\rho)$ remains monotonic for all values of $\rho$ once stringy corrections are taken into account and that $L(\rho)$ remains finite as $\rho$ approaches zero signaling pair creation.

\subsection{Beta Functions}

\begin{figure}[!h]
\begin{center}
\includegraphics[width=4.5in]{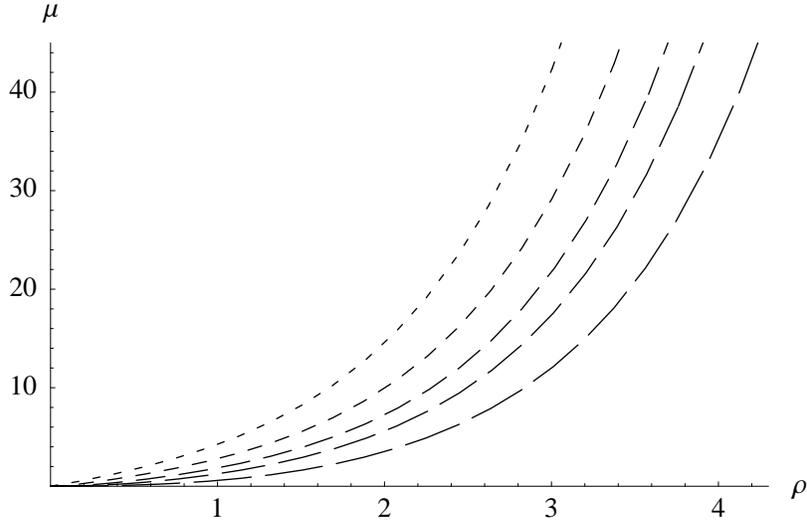}
\caption{This plot shows the Wilsonian cut-off as function of $\rho$ for $\rho_0=0.1, 0.5, 1, 1.4, 2$. $\mu$ is given in units of $\frac{e^{\phi_0} \sqrt{g_s N_c}}{2 \pi \sqrt{\alpha'}}$ and we have dashed the lines so that longer dashes correspond to larger $\rho_0$ values.} \label{fig:mus}
\end{center}
\end{figure}

Recall that the Wilsonian beta function for SQCD is given by
\begin{equation}
\beta_g = -\frac{g^3_{SQCD}}{16 \pi^2} (3 N_c - N_f (1-\gamma_0)),
\end{equation}
where $\gamma_0$ is the anomalous dimension of the quark superfield. We choose to calculate the equivalent quantity
\begin{align}\label{eq:betafunction}
\beta = \mu \frac{d}{d\mu} \frac{4 \pi^2}{\lambda_c}= \frac{1}{2 N_c} (3 N_c -N_f (1-\gamma_0)).
\end{align}
There are two distinct ways to establish a relation between the Wilsonian cut-off $\mu$ and the radial direction $\rho$ discussed in the literature \cite{Peet:1998wn}, sometimes referred to as the holographic relation and the stretched string relation. Since we only know our solutions numerically, and the stretched string relation is much easier to implement, we will restrict ourselves to the latter. In other words we use the following relation between the Wilsonian cut-off and the radial distance
\begin{equation}
\mu(\rho)= \frac{\sqrt{g_s}}{2  \pi \sqrt{\alpha'}}\int_0^{\rho} e^{\phi(\rho')+k(\rho')}d\rho'.
\end{equation}
This is shown for our usual representatives in Figure \ref{fig:mus}.
\begin{figure}[!h]
\begin{center}
\includegraphics[width=4.5in]{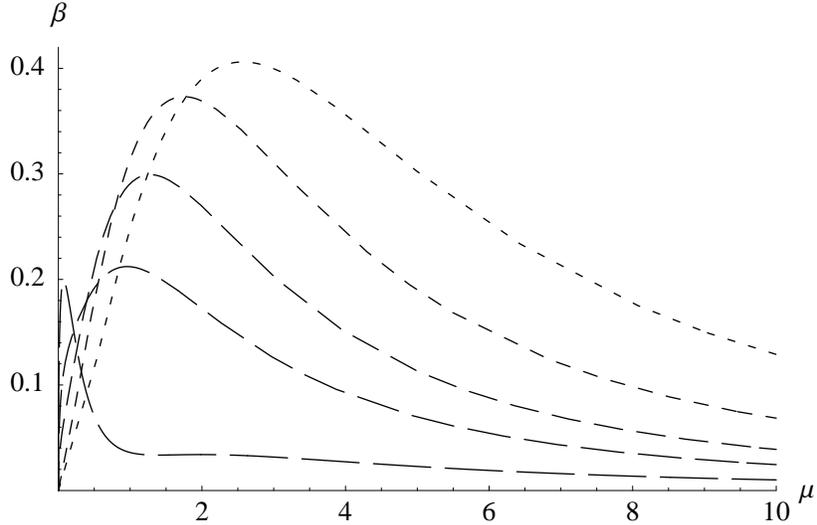}
\caption{This plot shows the beta function \protect\eqref{eq:betafunction} for $\rho_0=0.1, 0.5, 1, 1.4, 2$. $\mu$ is given in units of $\frac{e^{\phi_0} \sqrt{g_s N_c}}{2 \pi \sqrt{\alpha'}}$ and we have dashed the lines so that longer dashes correspond to larger $\rho_0$ values.} \label{fig:betafunctions}
\end{center}
\end{figure}\\
The range of $1.2 \lesssim \rho \lesssim 2.5$ for which both the supergravity approximation is valid and the KK-modes are negligible corresponds to one order of magnitude in energy for the solution starting at $e^{g+h}/N_c=\rho_0 =1.4$ and $e^{2k}=0$. The range can be made arbitrarily large by making $\rho_0$ small.\\
In Figure \ref{fig:betafunctions} we show the numerical results for the beta functions. For the range of $1.2 \lesssim \rho \lesssim 2.5$ the beta functions are monotonically decreasing. They start at values smaller than $0.5$ and approach zero for large values of $\mu$. This means that the anomalous dimension is always negative and approaches $-1/2$ from above in the UV.

\section{Conclusions and Future Directions}\label{sec:concl}
In this paper we found new supergravity backgrounds incorporating the backreaction of a large number $N_f=2N_c$ of flavor D5-branes. Among the solutions presented we identify a one-parameter family of solutions that describes the corresponding near-horizon geometries. We therefore expect this family to be dual to an SQCD-like theory. We calculated the Wilson loops and beta functions of these backgrounds finding a behavior that is consistent with the claim that it is  dual to an ${\mathcal N}=1$ SQCD-like theory  with $N_f=2N_c$ matter in the fundamental representation. 
There are several issues related to the present work that deserve further study:

\begin{itemize}
\item The field theory dual to the backgrounds presented here needs to be better understood. This problem can be approached from different, complimentary, angles. From the supergravity side it would be interesting to calculate field theory observables like spectra of glueballs, mesons and  baryons and compare them with results for the CVMN background \cite{Nunez:2003cf},\cite{Caceres:2005yx}. Seiberg duality and its geometric realization needs to be explored. It would also be very desirable to have a detailed construction of the field theory from the open string side. 

\item  The universality of the shear viscosity to entropy density coefficient  \cite{Policastro:2001yc},\cite{Policastro:2002se},\cite{Kovtun:2004de} was a successful result that opened the possibility of studying real life finite temperature QCD with gauge/gravity methods. Several QGP signatures have been investigated \cite{Herzog:2006gh} yielding a promising picture. However, all the non-extremal gravity backgrounds
used to date do not incorporate dynamical quarks. It is unclear if finite temperature  $\mathcal N =4$ SYM can capture the dynamics  of QCD above the deconfinement transition with enough precision to be useful in RHIC physics. Thus, finding a black hole solution of the backgrounds presented here and studying its hydrodynamic properties as well as QGP signatures is undoubtedly an interesting problem but we suspect that one might find problems similar to the ones encountered in \cite{Gubser:2001eg}. There it was observed that in the CVMN background the SYM modes do not seem to cleanly decouple from the massive Kaluza-Klein modes and there seem to be no black hole solutions for which the Hawking temperature is less than the Hagedorn temperature of LST. It still seems worthwhile to understand whether these problems persist in our setup.

\item Another avenue to explore in the quest for a QCD-like dual is finding a  stable non-SUSY deformation \cite{Aharony:2002vp} of the backgrounds presented here.
\end{itemize}

\section{Acknowledgments}
It is a pleasure  to thank Carlos Nu{\~n}ez  for many  useful discussions, comments   and correspondence. We are very grateful   to Arkady A. Tseytlin for reading the draft as well as for his valuable comments. We also  thank Francesco Benini, Alex Buchel, Jacques Distler, Hans Jockers, Daniel Robbins and Sonia Paban for helpful discussions. Elena C\'aceres thanks the Theory Group at the University  of Texas at Austin for hospitality during the completion of this work. The research of the authors is based upon work supported by the National Science Foundation under Grant No. PHY-0455649. The research of E.C is also funded by CONACYT grant 50760.

\appendix

\section{Appendix: Stability Analysis}\label{ap:stability}
In this appendix we will discuss the stability of probe D5-branes in both the CVMN background and our background that takes the backreaction to the presence of smeared flavor branes into account. The probes will extend along the four-dimensional spacetime, along the Hopf fiber inside the three-sphere parametrized by $\psi$, and along the radial direction. While Euler angles are convenient coordinates on the three-sphere in the rest of our paper, they are not very convenient for the stability analysis and we change to spherical polar coordinates on the $S^3$ as follows
\begin{subequations}
\begin{eqnarray}
X^4=\cos\frac{\tilde\theta}{2}\cos\frac{\tilde\varphi+\psi}{2}&=&\sin w\cos v\,,\\
X^3=\cos\frac{\tilde\theta}{2}\sin\frac{\tilde\varphi+\psi}{2}&=&\cos w\,,\\
X^2=\sin\frac{\tilde\theta}{2}\sin\frac{\tilde\varphi-\psi}{2}&=&-\sin w\sin v\sin u\,,\\
X^1=\sin\frac{\tilde\theta}{2}\cos\frac{\tilde\varphi-\psi}{2}&=&\sin w\sin v\cos u\,.
\end{eqnarray}
\end{subequations}
If we denote the worldvolume coordinates by $(\xi^\mu,\sigma,\rho)$,\footnote{We take $\sigma$ to have period $2\pi$.} the cylinder solution we are interested in is described by the embedding
\begin{equation}
\begin{array}{ll}
x^\mu(\xi^\mu,\sigma,\rho) = \xi^\mu\,, & u(\xi^\mu,\sigma,\rho) = \sigma\,,  \\
r(\xi^\mu,\sigma,\rho) = \rho\,, & v(\xi^\mu,\sigma,\rho) = \frac{\pi}{2}\,, \\
\theta(\xi^\mu,\sigma,\rho) = \theta_0\,, & w(\xi^\mu,\sigma,\rho) = \frac{\pi}{2}\,, \\
\varphi(\xi^\mu,\sigma,\rho) = \varphi_0.
\end{array}
\end{equation}
Using reparametrization invariance of the worldvolume theory, we can write the most general fluctuations around this background as follows
\begin{equation}
\begin{array}{ll}
x^\mu(\xi^\mu,\sigma,\rho) = \xi^\mu\,, & u(\xi^\mu,\sigma,\rho) = \sigma\,,  \\
r(\xi^\mu,\sigma,\rho) = \rho\,, & v(\xi^\mu,\sigma,\rho) = \frac{\pi}{2} +\alpha(\xi^\mu,\sigma,\rho)\,,\\
\theta(\xi^\mu,\sigma,\rho) = \theta_0 +\tilde\alpha(\xi^\mu,\sigma,\rho)\,,& w(\xi^\mu,\sigma,\rho) = \frac{\pi}{2}+\beta(\xi^\mu,\sigma,\rho)\,, \\
\varphi(\xi^\mu,\sigma,\rho) = \varphi_0+\tilde\beta(\xi^\mu,\sigma,\rho).
\end{array}
\end{equation}
The fluctuations $\tilde\alpha, \tilde\beta$ correspond to translations on the two-sphere and it is easy to see that their zero-modes will have a flat potential allowing us to drop them from our analysis. Furthermore, we can convince ourselves that $\sigma$ and $\xi^i$ dependence of $\alpha$ and $\beta$ will yield a positive contribution to their energy. Therefore, when analyzing the stability, we can limit ourselves to fluctuations of the form
\begin{equation}
\begin{array}{ll}
x^\mu(\xi^\mu,\sigma,\rho) = \xi^\mu, & u(\xi^\mu,\sigma,\rho) = \sigma\,,  \\
r(\xi^\mu,\sigma,\rho) = \rho\,, & v(\xi^\mu,\sigma,\rho) = \frac{\pi}{2} +\alpha(t,\rho)\,,\\
\theta(\xi^\mu,\sigma,\rho) = \theta_0\,, & w(\xi^\mu,\sigma,\rho) = \frac{\pi}{2}+\beta(t,\rho)\,, \\
\varphi(\xi^\mu,\sigma,\rho) = \varphi_0.
\end{array}
\end{equation}
 
\subsection{Stability of the Cylinder Solution in the CVMN Background}
\begin{figure}[!h]
\begin{center}
\includegraphics[width=4.2in]{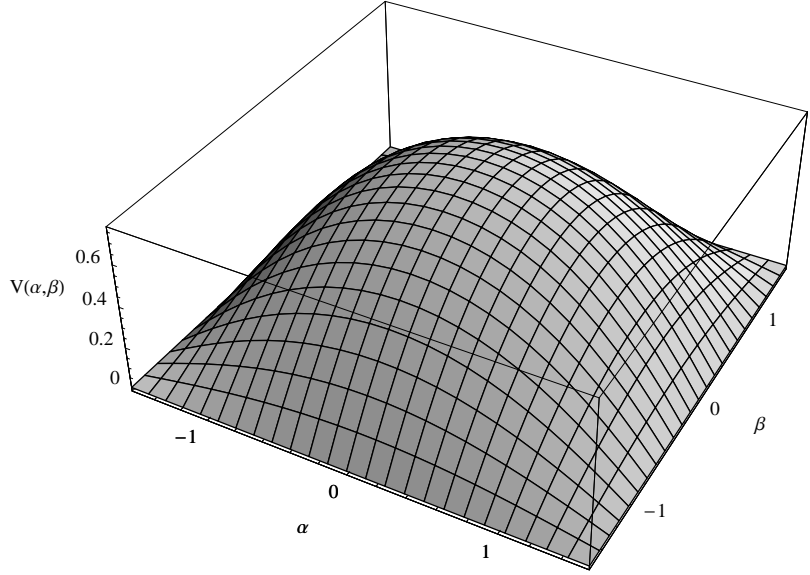}
\end{center}
\caption{This plot shows the potential for fluctuations around the cylinder solution in the CVMN background for a generic value of the radial variable.} \label{fig:mnpotential}
\end{figure}

\begin{figure}[!h]
\begin{center}
\includegraphics[width=4in]{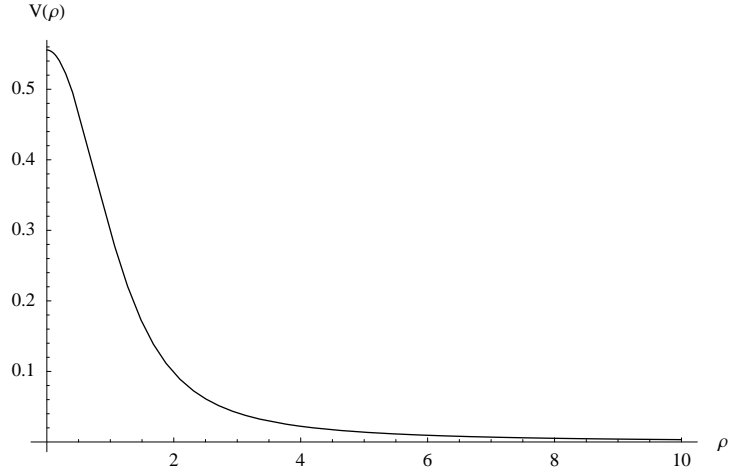}
\end{center}
\caption{This plot shows the potential \eqref{eq:mnpotential} of the Schr{\"o}dinger problem \eqref{eq:mnsch}.} 
\label{fig:mnschpot}
\end{figure}

We are now ready to study whether or not there are unstable modes corresponding to the brane slipping off the three-sphere.
The worldvolume action is given by
\begin{equation}\label{eq:probeaction}
S_{\text{flavor}}= -T_5 \int_{\MM_6}d^6x e^{\phi/2} \sqrt{-g_{(6)}} + T_5\int_{\MM_6} P[C_{(6)}],
\end{equation}
where $g_{(6)}$ as usual is the determinant of the pull-back of the metric on the target space and $P[C_{(6)}]$ denotes the pull-back of the six-form via the embedding map. The six-form is given by
\begin{equation}
C_{(6)}=\frac{e^{2\phi}}{8}dx^0\wedge dx^1\wedge dx^2\wedge dx^3 \wedge\mathcal{C}\,,
\end{equation} 
where $\mathcal{C}$ is given by
\begin{multline}
\mathcal{C}=\left((a^2-1)a^2e^{-2h}-16e^{2h}\right)\cos\theta d\varphi\wedge dr\\-(a^2-1)e^{-2h}\omega^3\wedge dr+a'(\sin\theta d\varphi\wedge\omega^1+d\theta\wedge\omega^2)\,.
\end{multline} 
For our embedding we find that, in units of the brane tension, this leads to a potential for the fluctuations given by
\begin{equation}
V(\alpha,\beta)= e^{2\phi}\left(\cos\alpha\cos\beta-\frac14(1-a^2)e^{-2h}\cos^2\alpha\cos^2\beta\right).
\end{equation} 
This depends on the radial coordinate via the dilaton as well as the functions $a$ and $h$. It turns out to be tachyonic for all values of the radial variable. It is shown for a sample value of the radial variable in Figure \ref{fig:mnpotential}.
\\
To quadratic order in fluctuations the action \eqref{eq:probeaction} takes the form\footnote{A prime denotes a derivative with respect to the radial variable.}
\begin{multline}
S_{\text{flavor}}= T_5 \int_{\MM_6}d^6x e^{2\phi} \left(\frac12(\dot\alpha^2+\dot\beta^2)-\frac12({\alpha'}^2+{\beta'}^2)\right.\\\left.-\frac12\left(\frac12e^{-2h}(1-a^2)-1\right)\left(\alpha^2+\beta^2\right)\right).
\end{multline}
Clearly this quadratic action has an $SO(2)$ symmetry acting on the fluctuations $\alpha$ and $\beta$, so that only $\delta(t,\rho)=\sqrt{\alpha(t,\rho)^2+\beta(t,\rho)^2}$ has a potential and we can focus our attention on this mode. Its equation of motion takes the form
\begin{equation}
\ddot\delta=\delta''+2\phi'\delta'+(1-\frac12e^{-2h}(1-a^2))\delta\,.
\end{equation}
As usual, this can be solved by separation of variables. Introducing $\delta(t,\rho)=T(t)e^{-\phi}\Psi(\rho)$ we find the following system of ordinary differential equations
\begin{eqnarray}
&\ddot T=-M^2T\,,\\
&-\Psi''+\left({\phi'}^2+\phi''-1+\frac12e^{-2h}(1-a^2)\right)\Psi=M^2\Psi\,.\label{eq:mnsch}
\end{eqnarray}
We see that there will be a tachyon, {\it i.e.} an exponentially growing mode, if the second equation has normalizable solutions with negative eigenvalues.\footnote{The condition of normalizability is imposed on us from the requirement that the energy in a fluctuation should be finite.} 
This is now just a one-dimensional quantum mechanics exercise with a potential 
\begin{equation}\label{eq:mnpotential}
V(\rho)={\phi'}^2+\phi''-1+\frac12e^{-2h}(1-a^2)=\frac{5(\cosh4\rho-8\rho^2-1)^2}{(\cosh4\rho+8\rho^2-1-4\rho\sinh4\rho)^2}\,,
\end{equation} 
which is shown in Figure \ref{fig:mnschpot}.\\
\\
Clearly, since the potential is positive definite, there are no eigenfunctions with negative eigenvalues and hence no tachyons implying that the cylinder solution is indeed stable in the CVMN background. 
In fact, using the usual norm we are used to from quantum mechanics,\footnote{To analyze stability in gravitational systems with a timelike Killing vector one checks that the energy is positive for all fluctuations for which it converges. The usual quantum mechanics norm in our case is equivalent to the norm induced by the energy functional.} there are no normalizable eigenfunctions at all, implying a continuous spectrum of fluctuations. This might be somewhat surprising since one might expect these fluctuations to correspond to part of the meson spectrum. We might wonder if one should in fact cut off the integral at the value of the radial variable at which the dilaton becomes too large to be able to trust supergravity, which would make the spectrum discrete. 
Fortunately, independent of this issue, the positivity of the potential already implies stability. Together with kappa symmetry this tells us that the cylinder solution is indeed supersymmetric in the CVMN background and it seems reasonable to smear it over the transverse directions. One might be worried that the smearing could introduce an additional instability but we will see in the next subsection that this is not the case. We attribute this to the fact that the fiber the branes wrap is non-trivially fibered over the base space.\\
We should point out that we have chosen the orientation of the brane so that it is kappa symmetric as well. If one reverses the orientation of the brane the last term in the potential changes sign and the potential is in fact no longer positive definite and does admit a bound state with negative energy corresponding to a tachyon. 

\subsection{Stability of the Cylinder Solution in the Backreacted Background}
\begin{figure}[!h]
\begin{center}
\includegraphics[width=4.2in]{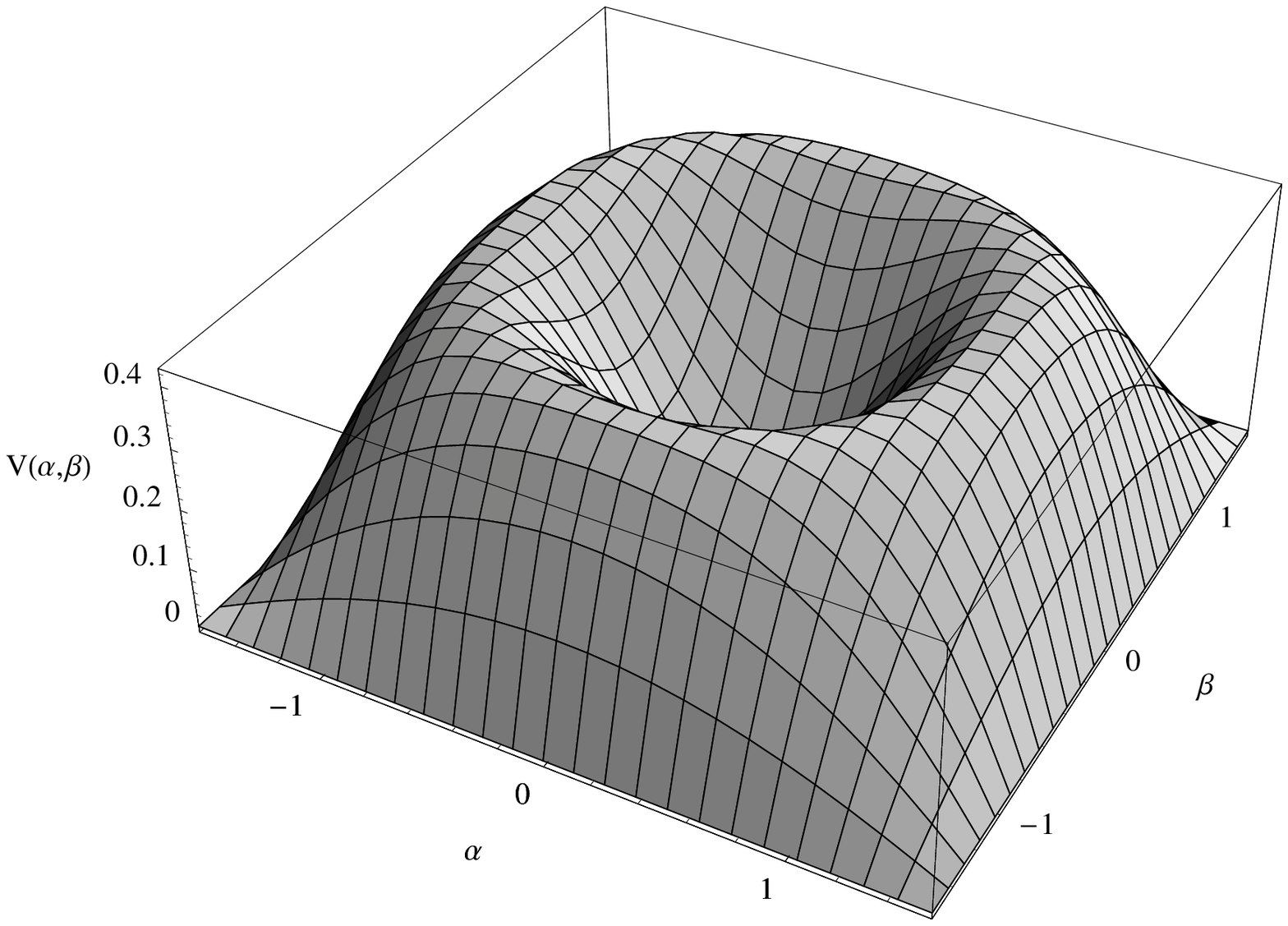}
\end{center}
\caption{This plot shows the potential for fluctuations around the cylinder solution in the backreacted background for a generic value of the radial variable.} \label{fig:2potential}
\end{figure}

\begin{figure}[!h]
\begin{center}
\includegraphics[width=4in]{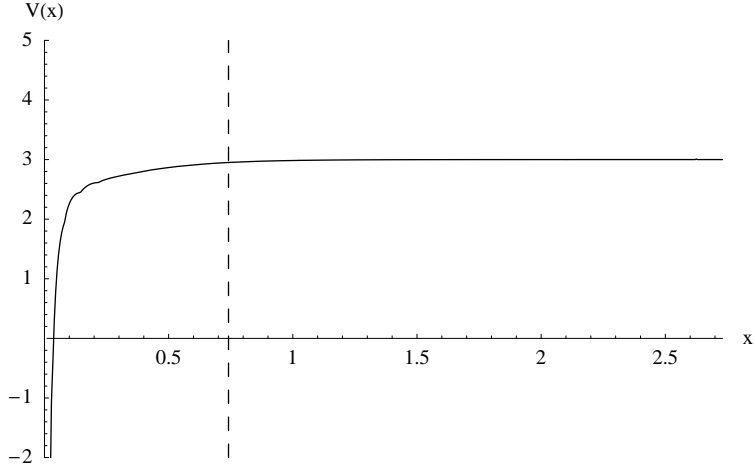}
\end{center}
\caption{This plot shows the potential \eqref{eq:2potential} of the Schr{\"o}dinger problem \eqref{eq:2sch} for the near horizon geometry with $\rho_0=1$.} \label{fig:2schpot}
\end{figure}

Now that we have convinced ourselves that a single probe brane in the CVMN background embedded as described above is supersymmetric it makes sense to try to smear it and take the backreaction into account. To make sure that the smearing does not spoil the supersymmetry we check that an additional probe D5-brane is still stable. In Appendix \ref{ap:BPS-eqs} we check that it is also kappa symmetric.
Again we will see that the only possibly dangerous modes are the modes corresponding to the brane slipping off the three-sphere.
The worldvolume action is of course still given by \eqref{eq:probeaction}. The six-form in our case is given by
\begin{equation}
C_{(6)}=\frac{N_c}{2}e^{2\phi}dx^0\wedge dx^1\wedge dx^2\wedge dx^3 \wedge d\rho\wedge (\omega^3-A^3)\,.
\end{equation} 
For our embedding we find that this leads to a potential for the fluctuations given by
\begin{multline}
V(\alpha,\beta)=e^{2\phi}\left(e^k\cos\alpha\cos\beta\sqrt{e^{2g}(\sin^2\alpha\cos^2\beta+\sin^2\beta)+e^{2k}\cos^2\alpha\cos^2\beta}\right.\\\left.\vphantom{e^k\cos\alpha\cos\beta\sqrt{e^{2g}(\sin^2\alpha\cos^2\beta+\sin^2\beta)}}-N_c\cos^2\alpha\cos^2\beta\vphantom{\sqrt{e^k}(a)}\right).
\end{multline} 
Again this depends on the radial coordinate via the dilaton as well as through the functions $k$ and $g$. For all values of the radial variable for which the supergravity approximation can be trusted this is no longer tachyonic. The presence of the flavor branes has stabilized this mode.\footnote{This statement clearly depends on the orientation of the probe brane. We have checked that for the kappa symmetric configuration it is stabilized while for the opposite orientation it is tachyonic.} It is shown for a sample value of the radial variable for the solution in the decoupling limit with $\rho_0=1$ in Figure \ref{fig:2potential}.
\\
From this we can already guess that the configuration will be stable but we will be careful and complete our analysis.
To quadratic order in fluctuations the action \eqref{eq:probeaction} takes the form\footnote{A prime denotes a derivative with respect to the radial variable, $\rho$.}
\begin{multline}
S_{\text{flavor}}= T_5 \int_{\MM_6}d^6x e^{2\phi} \left(\frac12e^{2k+2g}(\dot\alpha^2+\dot\beta^2)-\frac12e^{2g}({\alpha'}^2+{\beta'}^2)\right.\\\left.-\frac12\left(e^{2g}-2(e^{2k}-N_c)\right)\left(\alpha^2+\beta^2\right)\right).
\end{multline}
Again, this action has an $SO(2)$ symmetry acting on the fluctuations $\alpha$ and $\beta$, so that only $\delta(t,\rho)=\sqrt{\alpha(t,\rho)^2+\beta(t,\rho)^2}$ has a potential and again we focus our attention on this mode. Its equation of motion takes the form
\begin{equation}
\ddot\delta=e^{-2\phi-2k-2g}\left(e^{2\phi+2g}\delta'\right)'-e^{-2k-2g}(e^{2g}-2(e^{2k}-N_c))\delta\,.
\end{equation}
Introducing $\delta(t,\rho(x))=T(t)e^{-\phi-g-k/2}\Psi(x)$ and defining a new radial coordinate $x$ via $dx=e^{k}d\rho$ we find the following system of ordinary differential equations
\begin{eqnarray}
&\ddot T=-M^2T\,,\\
&-\Psi''+\left((\frac{k'}{2}+g'+\phi')^2+(\frac{k''}{2}+g''+\phi'')+e^{-2k-2g}(e^{2g}-2(e^{2k}-N_c)) \right)\Psi=M^2\Psi,\label{eq:2sch}
\end{eqnarray}
where the prime now indicates a derivative with respect to $x$.
As before a tachyon corresponds to a normalizable solution of the second equation with negative eigenvalue. 
Again we have reduced the problem to a one-dimensional quantum mechanics exercise with a potential 
\begin{equation}\label{eq:2potential}
V(x)=\left(\frac{k'}{2}+g'+\phi'\right)^2+\left(\frac{k''}{2}+g''+\phi''\right)+e^{-2k-2g}(e^{2g}-2(e^{2k}-N_c))\,,
\end{equation}
which is shown in Figure \ref{fig:2schpot}.
\\
Somewhat surprisingly, the potential is no longer positive definite and it even seems that it might admit bound states with negative energy eigenvalues. However, we have to remember that our supergravity solution is singular and can only be trusted far enough from the singularity. So we argue that one should impose a hard cutoff, {\it i.e.} introduce an infinitely high wall, at the value of the radial variable below which supergravity breaks down. The value of the radial variable below which supergravity breaks down for this particular case is indicated by the dashed line. With a cutoff the  potential will once again not lead to any eigenfunctions with negative eigenvalues and we conclude that tachyons are absent even once we have smeared the branes and taken the backreaction into account. The same remarks about the norm that we made for the CVMN background apply for our solutions as well but this issue once again does not alter the conclusion that the configuration is stable under small fluctuations.

\section{Appendix: Derivation of BPS Equations and Kappa Symmetry}\label{ap:BPS-eqs}
In this appendix we derive the BPS equations of type IIB supergravity for our ansatz following \cite{Nunez:2003cf},\cite{Casero:2006pt}.\footnote{We are considering the special case $b(\rho)=0, x=2$ of \cite{Casero:2006pt} and use the rescaled radial coordinate $d\rho= e^{-k}dr$.} We then derive the kappa symmetry condition \cite{Bergshoeff:1996tu},\cite{Bergshoeff:1998ha} for a single flavor probe brane localized in the $\theta, \varphi, \tilde \theta, \tilde \varphi$ directions. Since our background is a solution to the BPS equations, the stability of the branes shown in Appendix \ref{ap:stability} together with the kappa symmetry implies that the system is supersymmetric \cite{Bergshoeff:1997kr}. 

\subsection{BPS Equations} Recall that the RR 3-form field strength in the vielbein basis \eqref{eq:vielbein} is
\begin{align}
F_{(3)}&=- 2 N_c e^{-3f-2g-k} e^1 \wedge e^2 \wedge e^3 + \frac{N_c}{2} e^{-3f-2h-k} (a^2-1) e^\theta \wedge e^\varphi \wedge e^3\nonumber\\
&+ N_c e^{-3f-h-g-k} a (e^\theta \wedge e^2 +e^\varphi \wedge e^1) \wedge e^3.
\end{align}
The non-vanishing components of the spin connections are
\begin{align}
\omega^{x_i \rho}= e^{-f-k} f' e^{x_i}\,,&\quad\omega^{3\rho}=e^{-f-k}(f'+k')e^3\,,\nonumber\\
\omega^{\theta 1}= -\omega^{\varphi 2} =  \frac{1}{4} a' e^{-f-h+g-k} e^\rho\,,&\quad
\omega^{1 \varphi}= \omega^{2 \theta} = a \sinh (k-g) e^{-f-h} e^3 \,,\nonumber\\
\omega^{\theta \rho}=e^{-f-k} (f'+h') e^\theta + \frac14 a' e^{-f-h+g-k} e^1\,,&\quad
\omega^{\varphi \rho}=e^{-f-k} (f'+h') e^\varphi - \frac14 a' e^{-f-h+g-k} e^2\,,\nonumber\\
\omega^{1 \rho}=e^{-f-k} (f'+g') e^1 + \frac14 a' e^{-f-h+g-k} e^\theta\,,&\quad
\omega^{2 \rho}=e^{-f-k} (f'+g') e^2 - \frac14 a' e^{-f-h+g-k} e^\varphi\,,\nonumber\\
\omega^{23}=-e^{-f+k-2g} e^1 + a e^{-f-h} \cosh (k-g) e^\theta\,,&\quad \omega^{21}=(-e^{-f+k-2g}+2e^{-f-k})e^3-e^{-h-f} \cot\theta e^\varphi\,, \nonumber\\
\omega^{13}=e^{-f+k-2g} e^2 + a e^{-f-h} \cosh (k-g) e^\varphi\,,&\quad \omega^{\theta\varphi} = -\frac14 e^{-2h-f+k} (a^2-1) e^3 -e^{-h-f} \cot \theta e^\varphi\,,\nonumber
\end{align}
\vspace{-.8cm}
\begin{align}
\omega^{3\varphi}=&-\frac14 e^{-2h-f+k} (a^2-1) e^\theta+
ae^{-f-h} \sinh (k-g) e^1 \,,\nonumber\\
\omega^{3\theta}=&\frac14 e^{-2h-f+k} (a^2-1) e^\varphi+
ae^{-f-h} \sinh (k-g) e^2 \,.\nonumber
\end{align}
The supersymmetry variations for the dilatino $\lambda$ and the gravitino $\psi_\mu$ in the Einstein frame are given by \cite{Schwarz:1983qr}
\begin{align}
\delta \lambda =& \frac12 i\Gamma^\mu \epsilon^* \partial_\mu \phi+
\frac{1}{24}e^{\frac{\phi}{2}}\Gamma^{\mu_1\mu_2\mu_3}\,
\epsilon \,F_{\mu_1\mu_2\mu_3}\,,\nonumber\\
\delta \psi_\mu =& \partial_\mu \epsilon +\frac14 \omega_\mu^{ab}
\Gamma_{ab}\epsilon + \frac{e^{\frac{\phi}{2}}}{96}
(\Gamma_\mu^{\ \mu_1\mu_2\mu_3} - 9 \delta_\mu^{\mu_1}
\Gamma^{\mu_2\mu_3}) i \epsilon^* F_{\mu_1\mu_2\mu_3},
\end{align}
where $\lambda, \psi_\mu$ and $\epsilon$ are complex Weyl spinors.\\
If we write $\epsilon =\epsilon_+ + \epsilon_-$ such that $i \epsilon^* = \epsilon_+ - \epsilon_-$ and then solve the variations for $\epsilon_+$, it is easy to verify that this implies that they cannot be solved for $\epsilon_-$. So we can drop the index and impose that $i \epsilon^* =\epsilon$.\\
Furthermore, $\epsilon=\epsilon(\rho)$ due to the symmetries of our ansatz for the metric.\\
The spinor $\epsilon$ also has to satisfy the condition that
\begin{equation}
\Gamma_{\theta \varphi} \epsilon = \Gamma_{12} \epsilon.
\end{equation}
This follows for example from demanding that the $\theta$ dependent part of $\delta \psi_\varphi$ vanishes and corresponds to demanding that the color branes wrap a supersymmetric 2-cycle.\\
The dilatino variation is
\begin{align}
\delta \lambda &= \frac12 \phi' e^{-f} \Gamma_\rho \epsilon +\frac14 e^{\phi/2} \Big( -2 N_c e^{-3f-2g-k} \Gamma_{123} +\frac{N_c}{2} e^{-3f-2h-k} (a^2-1) \Gamma_{\theta \varphi 3} \nonumber\\
&  +N_c e^{-3f-h-g-k} a (\Gamma_{\theta 2} +\Gamma_{\varphi 1}) \Gamma_3 \Big)\epsilon.
\end{align}
Comparing this with the $x^\mu$ components of the gravitino variation we find that $\phi = 4f$. Using this and multiplying the equation above with $2 e^f \Gamma_{123}/\phi'$ we have
\begin{equation}\label{eq:projection}
\Gamma_{\rho 123} \epsilon = (\mathcal{A} + \mathcal B \Gamma_{\varphi 2}) \epsilon,
\end{equation}
where
\begin{align}
\mathcal A &= \frac{N_c}{\phi'} (e^{-2g} -\frac14 e^{-2h}(a^2-1)),\\
\mathcal B &= -\frac{N_c}{\phi'} a e^{-h-g}.
\end{align}
From the conditions $(\Gamma_{\rho123})^2 \epsilon =\epsilon$ and $\{\Gamma_{\rho123},\Gamma_{\varphi 2}\}=0$ it follows that \begin{equation}
\mathcal A^2+\mathcal B^2 =1.
\end{equation}
Therefore, we can write $\mathcal A=\cos \alpha, \, \mathcal B=\sin \alpha$. The projection condition \eqref{eq:projection} then becomes
\begin{equation}
\Gamma_{\rho 123} \epsilon = e^{\alpha \Gamma_{\varphi 2}} \epsilon.
\end{equation}
In terms of $\epsilon_0$ defined by $\epsilon = e^{-\frac{\alpha}{2} \Gamma_{\varphi 2}} \epsilon_0$ this projection conditions is simply
\begin{equation}
\Gamma_{\rho 123} \epsilon_0 = \epsilon_0.
\end{equation}
From the $\rho$ component of the gravitino variation we find
\begin{equation}
\partial_\rho \epsilon -\frac{\phi' \epsilon}{8} -\frac14 a' e^{-h+g} \Gamma_{\varphi 2} \epsilon=0,
\end{equation}
which gives
\begin{equation}
\partial_\rho \epsilon_0-\frac{\alpha'}{2} \Gamma_{\varphi 2} \epsilon_0 -\frac{\phi' \epsilon_0}{8} - \frac14 a' e^{-h+g} \Gamma_{\varphi 2} \epsilon_0 =0.
\end{equation}
Solving the terms with and without $\Gamma_{\varphi2}$ independently we have
\begin{equation}
\alpha'=-\frac12 a' e^{-h+g} \qquad \text{and} \qquad \epsilon_0 =e^{\phi/8} \eta,
\end{equation}
where $\eta$ is a constant spinor satisfying the same projections as $\epsilon_0$.\\ 
So to summarize, our Killing spinor $\epsilon$ is of the form
\begin{equation}
\epsilon=e^{-\frac{\alpha}{2}\Gamma_{\varphi2}}e^{\frac{\phi}{8}}\eta\,,
\end{equation}
where $\eta$ is a constant spinor satisfying the projections $$\Gamma_{\rho 1 2 3}\eta=\eta, \qquad \Gamma_{\theta\varphi}\eta=\Gamma_{12}\eta, \qquad i\eta^*=\eta $$ and $\alpha$ is obtained by integrating $$\alpha'=-\frac12 a' e^{-h+g}.$$
\\
From the remaining variations we obtain two equations each, one from the part proportional to $\epsilon$ and the other from terms proportional to $\Gamma_{\varphi 2} \epsilon$.\\
From $\delta\psi_{x^\mu}=0$ or $\delta \lambda =0$ we get
\begin{eqnarray}
f'&=&\frac{N_c}{4} {\mathcal A} e^{-2g}- \frac{N_c}{16}{\mathcal A} e^{-2h}
(a^2 -1)- \frac{1}{4} a {\mathcal B} e^{-h-g} N_c \,,\\
0 &=& \mathcal A a + \mathcal B e^{h-g} -\frac14 \mathcal B e^{g-h} (a^2-1).\label{eq:BPS1}
\end{eqnarray}
From $\delta\psi_\theta =0$ or equivalently from $\delta\psi_\varphi =0$
and using the previous expressions we get
\begin{eqnarray}
h'&=&-{\mathcal B} a e^{-h+k}\left(\cosh (k-g) - \frac{N_c}{2}
e^{-k-g} \right)-\frac14 {\mathcal A} e^{-2h + 2k} (a^2-1) \nonumber \\
&& + \frac{N_c}{4}{\mathcal A} e^{-2h} (a^2 -1)\,,\\
a'&=&-4 {\mathcal A} a e^{-g+k} \cosh (k-g) + {\mathcal B} e^{-g-h+2k} (a^2-1)
-2 N_c {\mathcal B} e^{h-3g} \nonumber \\
&& - \frac{N_c}{2}  {\mathcal B} e^{-h-g} (a^2 -1).
\end{eqnarray}
From $\delta \psi_1=0$ or  $\delta \psi_2=0$ we get
\begin{eqnarray}
g'&=&{\mathcal A} e^{2k-2g} - {\mathcal B} a e^{-h+k} \sinh (k-g)
+\frac{N_c}{2} {\mathcal B} a e^{-h-g} - N_c {\mathcal A}
e^{-2g}\,, \\
0&=&\mathcal A a + \mathcal B e^{h-g} -\frac14 \mathcal B e^{g-h} (a^2-1).\label{eq:BPS2}
\end{eqnarray}
Note that \eqref{eq:BPS2} is the same as \eqref{eq:BPS1}.\\
From $\delta \psi_3=0$ and \eqref{eq:BPS1} one finds
\begin{eqnarray}
k'&=&-{\mathcal A} e^{2k-2g}+2 {\mathcal A} + \frac14 {\mathcal A} e^{-2h+2k} (a^2 -1) + 2 a {\mathcal B} e^{-h+k} \sinh (k-g) \nonumber \\
&&- N_c {\mathcal A} e^{-2g}+ \frac{N_c}{4} {\mathcal A} e^{-2h} (a^2 -1) + N_c {\mathcal B} a e^{-h-g}\,,\\
0&=& 2{\mathcal B} +a {\mathcal A} e^{g-h}.
\end{eqnarray}
It is straightforward to check that for $a(\rho) = \frac{1}{\cosh{2\rho}}$ these equations imply
\begin{equation}
e^{h-g} = \frac12 \tanh{2 \rho}, \quad \mathcal A =\tanh{2 \rho}, \quad \mathcal B=-\frac{1}{\cosh{2 \rho}},
\end{equation}
and
\begin{align}
\partial_{\rho} e^{h+g} &= e^{2k} -N_c,\\
\partial_{\rho} k &= -(e^{2k}+N_c)e^{-h-g}+2 \coth 2\rho,\\
\partial_{\rho} f &= \frac{N_c}{4} e^{-g-h}.
\end{align}
\subsection{Kappa Symmetry} Now we follow \cite{Nunez:2003cf} and show that a single probe flavor brane in the backreacted background is kappa symmetric. We take the flavor brane to extend along the $x^\mu, \rho$ and $\psi$ directions and at fixed $\theta, \varphi, \tilde \theta, \tilde \varphi$. For $\epsilon= i \epsilon^*$ and in the absence of a worldvolume gauge field and $B$-field we have to show that
\begin{equation}
\Gamma_\kappa \epsilon =\epsilon,
\end{equation}
where
\begin{equation}
\Gamma_\kappa =\frac{1}{6!} \frac{1}{\sqrt{-g^{(6)}}} \epsilon^{m1 \ldots m6} \gamma_{m1 \ldots m6}.
\end{equation}
$g^{(6)}_{mn} =\partial_m X^\mu \partial_n X^\nu g_{\mu \nu}$ and $\gamma_m = \partial_m X^\mu e^a_\mu \Gamma_a$ are the pullback of the metric to the worldvolume of the brane and the induced Dirac matrices on the worldvolume of the brane, respectively. For our embedding we can choose $x^\mu, \psi$ and $\rho$ as worldvolume coordinates. This gives
\begin{equation}
\gamma_{x^\mu} = e^f \Gamma_{x^\mu}, \quad \gamma_\psi = \frac{e^{f+k}}{2} \Gamma_3, \quad \gamma_\rho = e^{f+k} \Gamma_\rho, \quad \sqrt{-g^{(6)}} = \frac12 e^{6f+2k}.
\end{equation}
So we need to show that
\begin{equation}
\Gamma_\kappa \epsilon = \Gamma_{x^0 x^1 x^2 x^3\,3 \, \rho} \epsilon=\epsilon.
\end{equation}
Since $\epsilon$ is a spinor of definite chirality of type IIB supergravity we know that it satisfies $\Gamma_{x^0 x^1 x^2 x^3 \rho \theta \varphi 123} \epsilon=\epsilon$. From multiplying \eqref{eq:projection} by $\Gamma_{x^0 x^1 x^2 x^3 \theta \varphi}$ we get
\begin{equation}
\Gamma_{x^0 x^1 x^2 x^3} (\cos \alpha \Gamma_{\theta \varphi} + \sin \alpha \Gamma_{\theta 2}) \epsilon = \epsilon,
\end{equation}
and from multiplying \eqref{eq:projection} with $\Gamma_{\theta \varphi}$ one can show that
\begin{equation}
(\cos \alpha \Gamma_{\theta \varphi} + \sin \alpha \Gamma_{\theta 2}) \epsilon =\Gamma_{3\rho} \epsilon.
\end{equation}
So we find that
\begin{equation}
\Gamma_\kappa \epsilon = \Gamma_{x^0 x^1 x^2 x^3 3\rho} \epsilon=\epsilon\,,
\end{equation}
which means that the probe brane is kappa symmetric. Note that this derivation does not depend on the angular coordinates and is therefore true independent of the brane's position.

\section{Appendix: Analytic Solutions}\label{ap:solutions}
In this appendix we derive the analytic solutions given above in Section \ref{subsec:sols}, compare them to the numerical solutions and discuss some of their properties.

\subsection{The Solution in Region I}
This region is characterized by $e^{2k}\ll N_c$. One might simply drop $e^{2k}$ from the equations \eqref{eq:diffeq2} but we will take a more systematic approach and treat $e^{2k}$ as a small perturbation. If we define
\begin{eqnarray}
e^{g+h}&=&N_c(\bar{\gamma}+\delta\gamma),\\
e^{2k}&=&N_c \delta k,
\end{eqnarray}
and keep only terms up to first order in small quantities, the equations become
\begin{equation}
\begin{split}
&\partial_\rho \bar\gamma=-1,\\
&\partial_\rho \delta\gamma=\delta k,\\
&\partial_\rho \ln\delta k=-\frac{2}{\bar\gamma}+4\coth 2\rho.
\end{split}
\end{equation}
This set of equations has the the solution 
\begin{subequations}\label{eq:solregIapprox}
\begin{align}
e^{g+h}/N_c=&\rho_0-\rho+\frac{1}{64{\rho_0}^2{\rho_I}^2}\left(\frac{1}{4}
\left(1+8(\rho-\rho_0)^2\sinh{4\rho}\right)\right.\\\nonumber
&-\left.(\rho-\rho_0)\cosh{4\rho}-\rho_0+\frac{8}{3}\rho(\rho^2-3\rho\rho_0 +3\rho_0^2)\right),\\
e^{2k}/N_c=&\left(\frac{\sinh{2\rho}}{2\rho_I}\left(1-\frac{\rho}{\rho_0} \right)\right)^2.
\end{align}
\end{subequations}
To zeroth order in $e^{2k}$ the result is
\begin{subequations}\label{eq:solregIzero}
\begin{align}
e^{g+h}/N_c&=\rho_0-\rho,\\
e^{2k}/N_c&=\left(\frac{\sinh{2\rho}}{2\rho_I}\left(1-\frac{\rho}{\rho_0} \right)\right)^2.
\end{align}
\end{subequations}

\begin{figure}[!h]
\begin{center}
\includegraphics[width=5in]{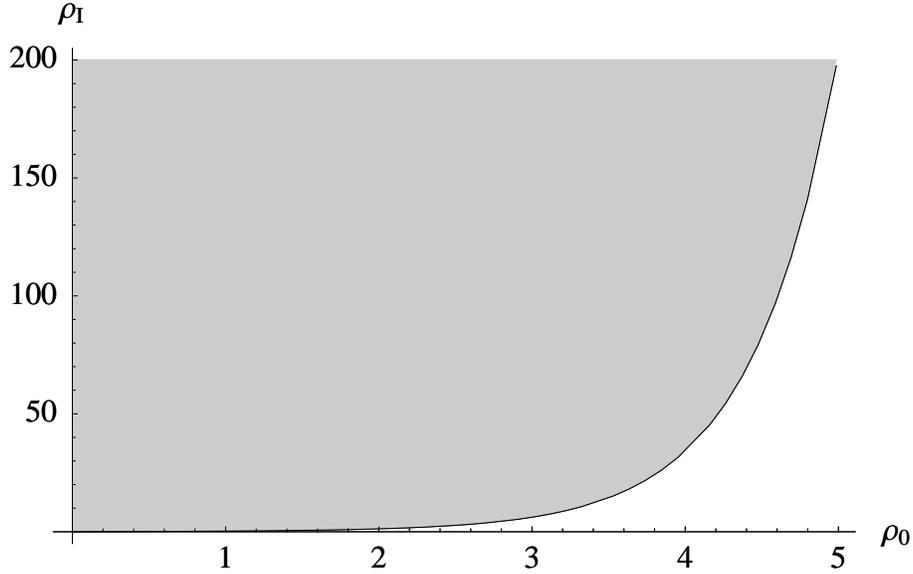}
\caption{The shaded region in this plot shows the region in $\rho_0 - \rho_I$ parameter space for which the solutions are contained entirely in Region I. The solid line  shows the function $\rho_{I,min}(\rho_0)$.}  \label{fig:contregI}
\end{center}
\end{figure}

{\bf Solutions Contained entirely in Region I}\\[.3cm]
In order for solutions to be contained entirely in Region I we have to impose that $e^{2k}\ll N_c$ for all values of $\rho$. This is satisfied if for all $\rho$
\begin{equation}\label{eq:contregI}
\frac{\sinh{2\rho}}{2\rho_I}\left(1-\frac{\rho}{\rho_0}\right)\ll1\,.
\end{equation}
For these solutions $e^{2k}$ has a maximum and it is enough to impose that the value at the maximum be much less than $N_c$. The maximum is located at a value $\overline\rho$ given by
\begin{equation}
0\stackrel{!}{=}\left.\frac{d}{d\rho}\right|_{\rho=\overline\rho}\left(\frac{
\sinh2\rho}{2\rho_I}\left(1-\frac{\rho}{\rho_0}\right)\right)=\frac{1}{
2\rho_0\rho_I}\left(2(\rho_0-\overline\rho)\cosh{2\overline\rho}-\sinh{
2\overline\rho}\right).
\end{equation}
So $\overline\rho$ is given by the solution of
\begin{equation}
2(\rho_0-\overline\rho)=\tanh{2\overline\rho}.
\end{equation}
From this it is clear, that $\overline\rho$ will only depend on $\rho_0$ and not on $\rho_I$ and we can rewrite the equation \eqref{eq:contregI} as a relation between $\rho_I$ and $\rho_0$
\begin{equation}\label{eq:rhoImin}
\rho_I\gg \frac{1}{2}\sinh{2\overline\rho(\rho_0)}\left(1-\frac{\overline\rho(\rho_0)}{\rho_0}\right) \equiv \rho_{I,min}(\rho_0).
\end{equation}
This can be solved numerically. The result is shown in Figure \ref{fig:contregI}.

\begin{figure}[!h]
\begin{center}
\includegraphics[width=6in]{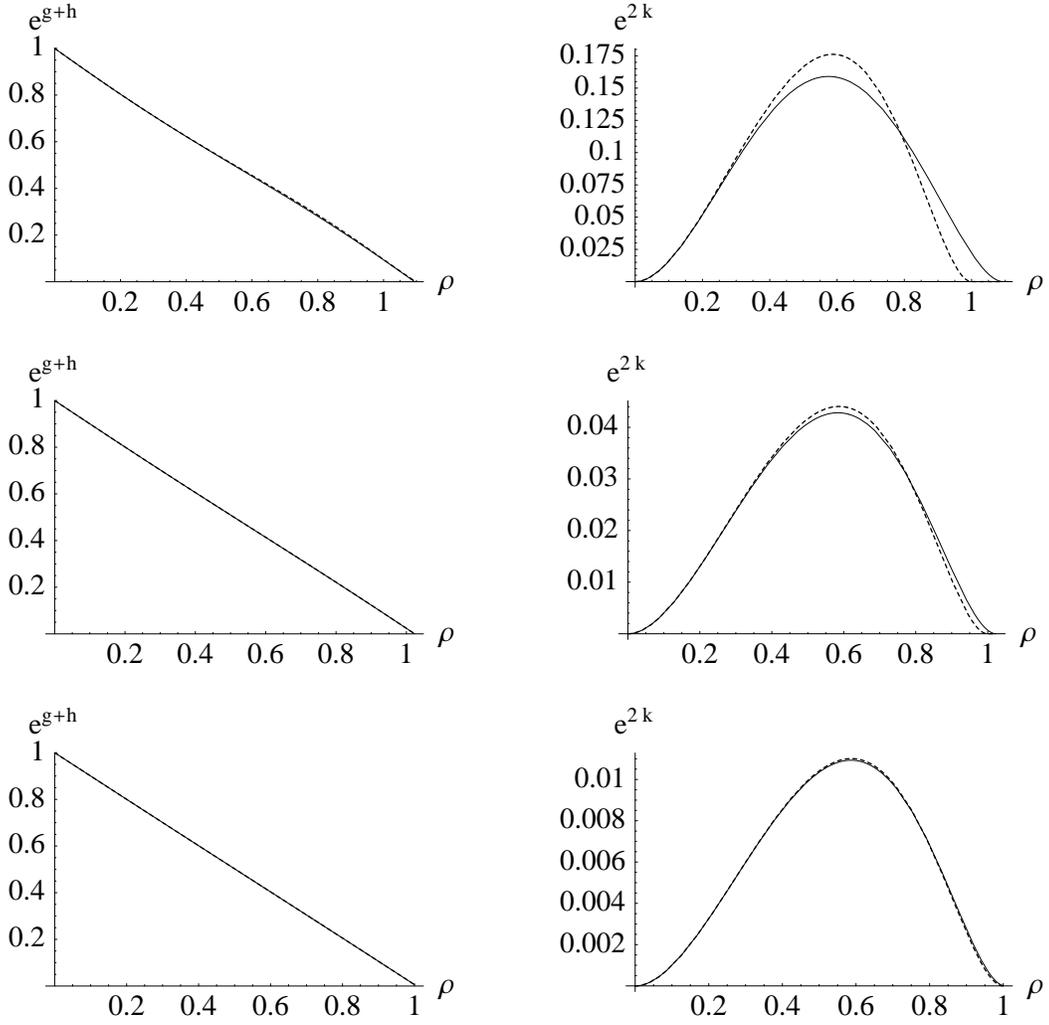}
\end{center}
\caption{Comparison of the exact numerical solution shown as a solid line and the approximate solution according to \protect\eqref{eq:solregIapprox} shown as a dashed line for $\rho_0=1$ and $\rho_I=2\rho_\text{I,min}(\rho_0)$, $\rho_I=4\rho_\text{I,min}(\rho_0)$, and $\rho_I=8\rho_\text{I,min}(\rho_0)$, in the top, middle, and bottom row, respectively. }
\label{fig:exvsappr1}
\end{figure}

\newpage
{\bf Comparison of Numerical and Analytic Solution in Region I}\\[.3cm]
To get an idea of how good the analytic solutions given in \eqref{eq:solregIapprox} are, we compare them to the exact numerical solutions for
$\rho_0=1$ in Figure~\ref{fig:exvsappr1} for three different values of $\rho_I$ given by 2, 4, and 8 times the bound $\rho_\text{I,min}$ according to \eqref{eq:rhoImin}. We see that the further we move inside the shaded region of Figure \ref{fig:contregI} the better the approximation becomes. We also see that for values of the parameters $\rho_0$ and $\rho_I$ near the border we should keep the next order in our expansion for $e^{2k}$ and that away from the border even the approximation \eqref{eq:solregIzero} is in excellent agreement with the numerical results.

\subsection{The Solution in Region II}
This region is characterized by $e^{g+h}\gg N_c,e^{2k}$. Under these assumptions the equations \eqref{eq:diffeq2} become
\begin{equation}
\begin{split}
&\partial_\rho e^{h+g}=e^{2k}-N_c,\\
&\partial_\rho k=2\coth 2\rho.
\end{split}
\end{equation}
These equations have the solution
\begin{eqnarray}
e^{g+h}/N_c&=&\rho_0-\rho-\frac{\rho}{8{\rho_I}^2}+\frac{\sinh{4\rho}}{32{\rho_I}^2},\\
e^{2k}/N_c&=&\left(\frac{\sinh{2\rho}}{2\rho_I}\right)^2.
\end{eqnarray}
Clearly these solutions will only be valid as long as
\begin{equation}
\rho_0-\rho-\frac{\rho}{8{\rho_I}^2}+\frac{\sinh{4\rho}}{32{\rho_I}^2}\gg 1, \, \left(\frac{\sinh{2\rho}}{2\rho_I}\right)^2.
\end{equation}

\subsection{The Solution in Region III}
This region is characterized by $e^{2k}\gg N_c$. The equations in this region become
\begin{equation}
\begin{split}
&\partial_\rho e^{h+g}=e^{2k},\\
&\partial_\rho e^{2k}=-2\frac{e^{4k}}{e^{g+h}}+4e^{2k}\coth 2\rho.
\end{split}
\end{equation}
These have the solution
\begin{subequations}\label{eq:solutionIII}
\begin{align}
e^{g+h}/N_c&=A\left(8B-12\rho+3\sinh{4\rho}\right)^{\frac{1}{3}},\\
e^{2k}/N_c&=\frac{8A\sinh^2{2\rho}}{(8B-12\rho+3\sinh{4\rho})^{\frac{2}{3}}}.
\end{align}
\end{subequations}
Here $A$ and $B$ are so far arbitrary integration constants. We will see how to relate them to more useful quantities below.\\

{\bf Solutions Contained entirely in Region III}\\[.3cm]
For solutions to be contained entirely in Region III the condition $e^{2k}\gg N_c$ has to hold for all $\rho$, in particular for $\rho\to0$. This implies that solutions contained entirely in Region III must have
\begin{equation}
B=0.
\end{equation}
So the solutions \eqref{eq:solutionIII} become
\begin{eqnarray}
e^{g+h}/N_c&=&3^{\frac{1}{3}}A\left(\sinh{4\rho}-4\rho\right)^{\frac{1}{3}},\\
e^{2k}/N_c&=&\frac{8A\sinh^2{2\rho}}{3^{\frac{2}{3}}(\sinh{4\rho}-4\rho)^{\frac{2}{3}}}.
\end{eqnarray}
After identifying
\begin{equation}
A=\frac{\epsilon^\frac{4}{3}}{2^\frac{4}{3}3^\frac{1}{3}}\,,
\end{equation}
these can be seen to correspond to the deformed conifold \cite{Candelas:1989js}
\begin{subequations}\label{eq:solregIIIapprox}
\begin{align}
e^{g+h}/N_c&=\frac{\epsilon^\frac{4}{3}\left(\sinh{4\rho}-4\rho\right)^{\frac{1}{3}}}{2^\frac{4}{3}},\\
e^{2k}/N_c&=\frac{2}{3}\epsilon^\frac{4}{3}\frac{2^\frac{2}{3}\sinh^2{2\rho}}{(\sinh{4\rho}-4\rho)^{\frac{2}{3}}}.
\end{align}
\end{subequations}
In order for this to be contained entirely in Region III we require
\begin{equation}
\epsilon\gg \left(\frac{3}{2}\right)^\frac{1}{4}\approx1\,.
\end{equation}

\begin{figure}[!h]
\begin{center}
\includegraphics[width=6in]{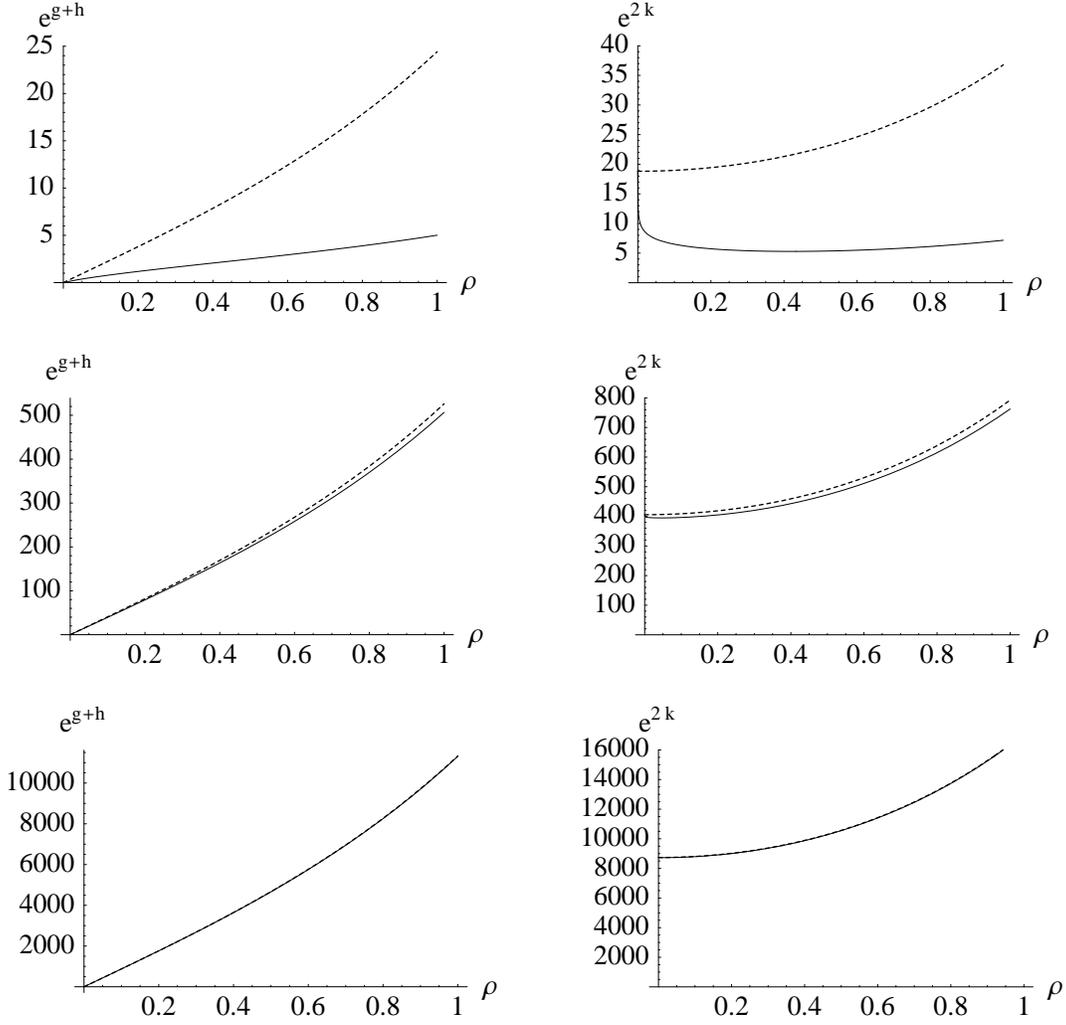}
\caption{Comparison of the exact numerical solution shown as a solid line and the approximate solution according to
\protect\eqref{eq:solregIIIapprox} shown as a dashed line for $\epsilon=10$, $\epsilon=100$, and $\epsilon=1000$, in
the top, middle, and bottom row, respectively.} \label{fig:exvsapprcon}
\end{center}
\end{figure}

{\bf Comparison of Numerical and Analytic Solution in Region III}\\[.3cm]
A comparison of these approximate solutions to the exact numerical solutions in this region are shown in Figure \ref{fig:exvsapprcon}. As was to be expected the approximate and numerical solution agree better and better the larger the deformation parameter $\epsilon$.

\subsection{Matching Solutions in Regions II and III}
The matching of the solutions in Regions II and III cannot be done in general but we can do so in two limiting cases, namely $\rho_I\ll1$ and $\rho_I\gg1$.\\

{\bf Matching for $\rho_I\ll1$}\\[.3cm]
In the case of small $\rho_I$ $e^{g+h}$ approaches its minimum value for small $\rho$ and for small enough $\rho_I$ the matching can be done for values of $\rho$ that are small compared to unity but large compared to $\rho_I$. For $\rho\ll1$ the solutions in Region II and III take the following form:\\

{\bf Region II}
\begin{eqnarray}
e^{g+h}/N_c&\to&\rho_0+\frac{\rho^3}{3{\rho_I}^2},\\
e^{2k}/N_c&\to&\frac{\rho^2}{{\rho_I}^2}.
\end{eqnarray}
In particular this implies that we need $\rho_0\gg1$ to ensure the validity of the approximations we used to find this solution.\\

{\bf Region III}
\begin{eqnarray}
e^{g+h}/N_c&\to&2A B^\frac{1}{3},\\
e^{2k}/N_c&\to&\frac{8A}{B^\frac{2}{3}}\rho^2.
\end{eqnarray}
This implies
\begin{equation}
A=\frac{1}{2}\left(\frac{\rho_0}{2\rho_I}\right)^\frac{2}{3}\text{  and  }B=4\rho_0\rho_I.
\end{equation}
So the solution can be written as
\begin{equation}
e^{g+h}/N_c=\left\{
\begin{array}{cc}
\rho_0-\rho-\frac{1}{8}\frac{\rho}{\rho_I^2}+\frac{\sinh{4\rho}}{32\rho_I^2}&\rho<\rho_*\\
\rho_0\left(1-\frac{1}{8\rho_0{\rho_I}^2}\left(3\rho-\frac{3}{4}\sinh{4\rho}\right)\right)^{\frac{1}{3}}&\rho>\rho_*
\end{array}
\right.,
\end{equation}
and similarly
\begin{equation}
e^{2k}/N_c=\left\{
\begin{array}{cc}
\left(\frac{\sinh{2\rho}}{2\rho_I}\right)^2&\rho<\rho_*\\
\left(\frac{\sinh{2\rho}}{2\rho_I}\right)^2\left(1-\frac{1}{8\rho_0{\rho_I}^2}\left(3\rho-\frac{3}{4}\sinh{4\rho}\right)\right)^{-\frac{2}{3}}&\rho>\rho_*
\end{array}
\right.,
\end{equation}
where $\rho_*$ is the value of $\rho$ for which we match the two solutions. For our approximations to be valid we require $\rho_0\gg1$ as well as $\rho_I\ll\rho_*\ll\sqrt{\rho_0}\rho_I$. We can choose a $\rho_*$ such that $\rho_*\ll\left(\rho_0\rho_I^2\right)^\frac{1}{3}$. Introducing a small additional error especially for small $\rho$ for $\rho_I\ll1$ we can write the solution as follows for all values of $\rho$
\begin{subequations}\label{eq:apprrIll1}
\begin{align}
e^{g+h}/N_c&=\rho_0\left(1-\frac{1}{8\rho_0 {\rho_I}^2} \left( 3 \rho- \frac{3}{4} \sinh{4\rho} \right) \right)^{\frac{1}{3}},\\
e^{2k}/N_c&=\left(\frac{\sinh{2\rho}}{2\rho_I}\right)^2\left(1-\frac{1}{8\rho_0{\rho_I}^2} \left( 3 \rho- \frac{3}{4} \sinh{4\rho} \right) \right)^{-\frac{2}{3}}.
\end{align}
\end{subequations}

{\bf Matching for $\rho_I\gg1$}\\[.3cm]
For $\rho_I\gg1$ the function $e^{g+h}$ takes its minimum for $\rho\gg1$ and so the matching will also be done for some value $\rho_*\gg1$. In this limit the solutions in Regions II and III take the form:\\

{\bf Region II}
\begin{eqnarray}
e^{g+h}/N_c&\to&\rho_0+\frac{e^{4\rho}}{64{\rho_I}^2},\\
e^{2k}/N_c&\to&\frac{e^{4\rho}}{16{\rho_I}^2}.
\end{eqnarray}
In particular this implies that we need $\rho_0\gg\frac{e^{4\rho_*}}{16\rho_I^2}\gg1$ to ensure the validity of the approximations we used to find this solution.\\

{\bf Region III}
\begin{eqnarray}
e^{g+h}/N_c&\to&A \left(8B+\frac{3}{2}e^{4\rho}\right)^\frac{1}{3},\\
e^{2k}/N_c&\to&\frac{2Ae^{4\rho}}{\left(8B+\frac{3}{2}e^{4\rho}\right)^\frac{2}{3}}.
\end{eqnarray}
Matching these then leaves us with the same conditions as before
\begin{equation}
A=\frac{1}{2}\left(\frac{\rho_0}{2\rho_I}\right)^\frac{2}{3}\text{  and  }B=4\rho_0\rho_I.
\end{equation}
So again the solution can be written as
\begin{equation}
e^{g+h}/N_c=\left\{
\begin{array}{cc}
\rho_0-\rho-\frac{1}{8}\frac{\rho}{\rho_I^2}+\frac{\sinh{4\rho}}{32\rho_I^2}&\rho<\rho_*\\
\rho_0\left(1-\frac{1}{8\rho_0{\rho_I}^2}\left(3\rho-\frac{3}{4}\sinh{4\rho}\right)\right)^{\frac{1}{3}}&\rho>\rho_*
\end{array}
\right.,
\end{equation}
and similarly
\begin{equation}
e^{2k}/N_c=\left\{
\begin{array}{cc}
\left(\frac{\sinh{2\rho}}{2\rho_I}\right)^2&\rho<\rho_*\\
\left(\frac{\sinh{2\rho}}{2\rho_I}\right)^2\left(1-\frac{1}{8\rho_0{\rho_I}^2}\left(3\rho-\frac{3}{4}\sinh{4\rho}\right)\right)^{-\frac{2}{3}}&\rho>\rho_*
\end{array}
\right.,
\end{equation}
where now the value of $\rho_*$ has to be chosen so that it satisfies
\begin{equation}
\frac{1}{2}\ln{4\rho_I}\ll\rho_*\ll\frac{1}{2}\ln{4\sqrt{\rho_0}\rho_I}.
\end{equation}
Clearly this is only possible if we take $\rho_0\gg1$ but this is consistent with what we have assumed above. If we take $\rho_0\gg\rho_*$ this simplifies further and we can, again introducing a small additional error, write the solution in closed form
\begin{subequations}\label{eq:apprrIgg1}
\begin{align}
e^{g+h}/N_c&=\rho_0\left(1+\frac{3e^{4\rho}}{64\rho_0{\rho_I}^2}\right)^{\frac{1}{3}}-\rho,\\
e^{2k}/N_c&=\left(\frac{\sinh{2\rho}}{2\rho_I}\right)^2 \left(1 + \frac{3e^{4\rho}}{64\rho_0{\rho_I}^2} \right)^{-\frac{2}{3}}.
\end{align}
\end{subequations}

\begin{figure}[!t]
\begin{center}
\includegraphics[width=6.5in]{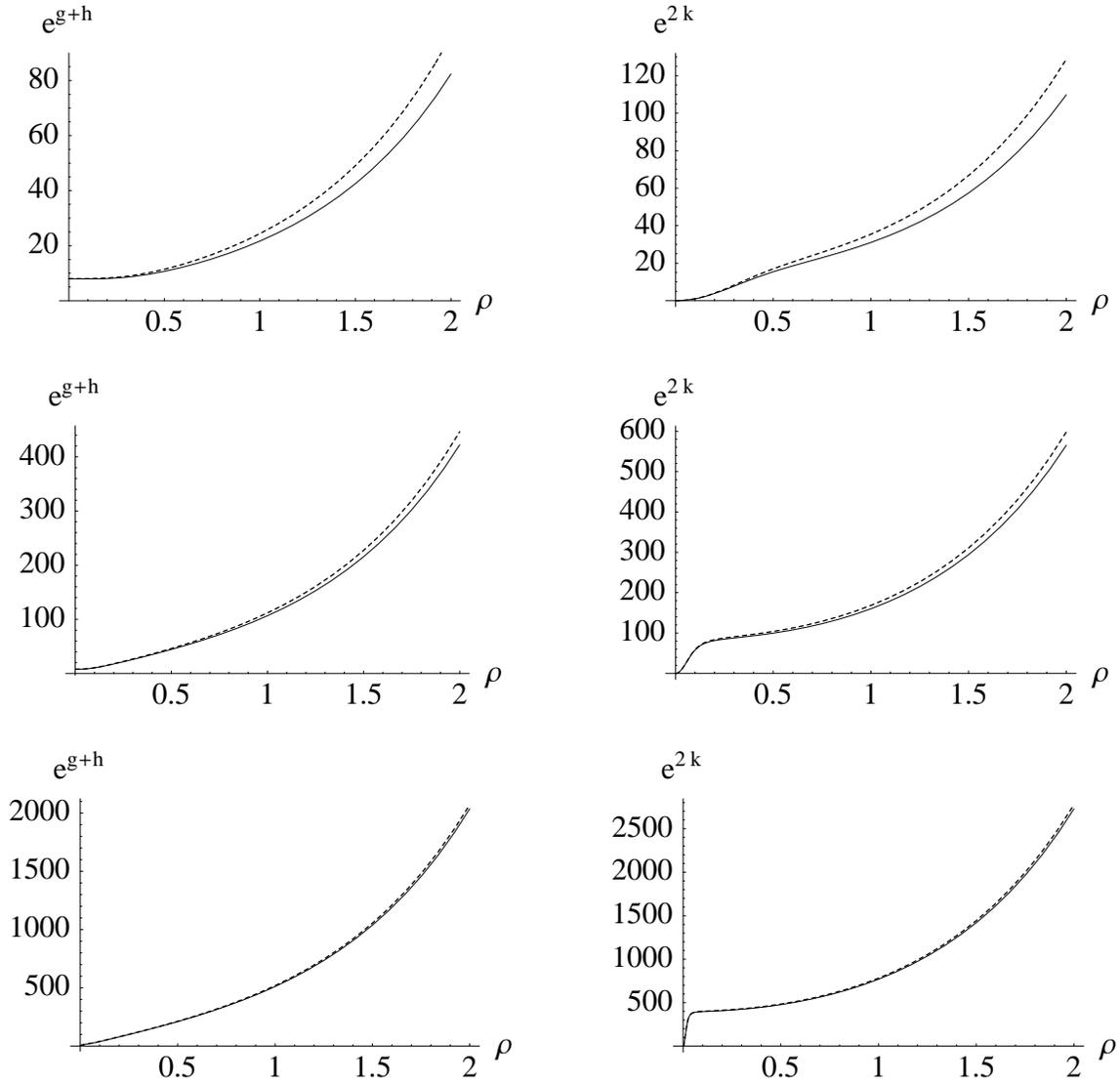}
\end{center}
\caption{Comparison of the exact numerical solution shown as a solid line and the approximate solution according to
\protect\eqref{eq:apprrIll1} shown as a dashed line for $\rho_0=8$ and $\rho_I=10^{-1}$, $\rho_I=10^{-2}$, and
$\rho_I=10^{-3}$, in the top, middle, and bottom row, respectively.} \label{fig:apprrIll1r08}
\end{figure}

\begin{figure}[!t]
\begin{center}
\includegraphics[width=6.5in]{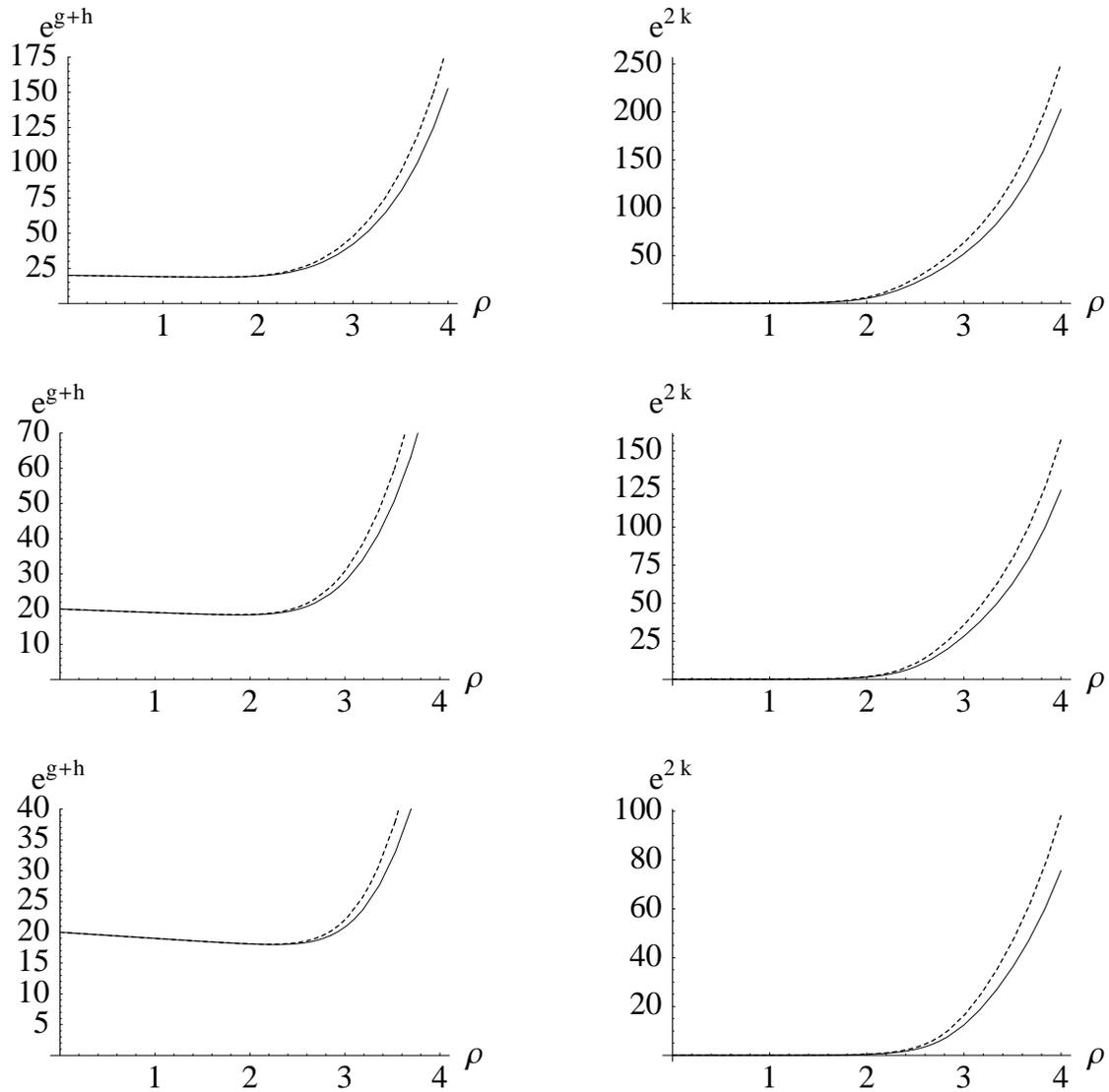}
\end{center}
\caption{Comparison of the exact numerical solution shown as a solid line and the approximate solution according to
\protect\eqref{eq:apprrIgg1} shown as a dashed line for $\rho_0=20$ and $\rho_I=5$, $\rho_I=10$, and $\rho_I=20$, in
the top, middle, and bottom row, respectively.} \label{fig:apprrIgg1r08}
\end{figure}

{\bf Comparison of Numerical and Analytic Solutions}\\[.3cm]
The approximation \eqref{eq:apprrIll1} is compared to the numerical solution for
$\rho_0=8$,
for $\rho_I=10^{-1}$, $\rho_I=10^{-2}$, and $\rho_I=10^{-3}$ in Figure
\ref{fig:apprrIll1r08}. As can be seen the approximation works reasonably well and the accuracy increases as $\rho_I$ decreases.
The approximation \eqref{eq:apprrIgg1} is compared to the numerical solution for $\rho_0=20$, and for $\rho_I=5$,
$\rho_I=10$, and $\rho_I=20$ in Figure~\ref{fig:apprrIgg1r08}. Again the approximation works reasonably well and the accuracy increases as $\rho_I$ increases.\newpage


\clearpage
\begin{thebibliography}{19}

\bibitem{Maldacena:1997re}
J.~M.~Maldacena,
``The large N limit of superconformal field theories and supergravity,''
Adv.\ Theor.\ Math.\ Phys.\  {\bf 2}, 231 (1998)
[Int.\ J.\ Theor.\ Phys.\  {\bf 38}, 1113 (1999)]
[arXiv:hep-th/9711200].

\bibitem{Witten:1998qj}
E.~Witten,
``Anti-de Sitter space and holography,''
Adv.\ Theor.\ Math.\ Phys.\  {\bf 2}, 253 (1998)
[arXiv:hep-th/9802150].

\bibitem{Aharony:1999ti}
O.~Aharony, S.~S.~Gubser, J.~M.~Maldacena, H.~Ooguri and Y.~Oz,
``Large N field theories, string theory and gravity,''
Phys.\ Rept.\  {\bf 323}, 183 (2000)
[arXiv:hep-th/9905111].

\bibitem{Chamseddine:1997nm}
A.~H.~Chamseddine and M.~S.~Volkov,
``Non-Abelian BPS monopoles in N = 4 gauged supergravity,''
Phys.\ Rev.\ Lett.\  {\bf 79}, 3343 (1997)
[arXiv:hep-th/9707176].

 
\bibitem{Chamseddine:1997mc}
A.~H.~Chamseddine and M.~S.~Volkov,
``Non-Abelian solitons in N = 4 gauged supergravity and leading order  string
theory,''
Phys.\ Rev.\  D {\bf 57}, 6242 (1998)
[arXiv:hep-th/9711181].



\bibitem{Maldacena:2000yy}
J.~M.~Maldacena and C.~Nunez,
``Towards the large N limit of pure N = 1 super Yang Mills,''
Phys.\ Rev.\ Lett.\  {\bf 86}, 588 (2001)
[arXiv:hep-th/0008001].
\bibitem{Klebanov:2000hb}
  I.~R.~Klebanov and M.~J.~Strassler,
  ``Supergravity and a confining gauge theory: Duality cascades and
  chiSB-resolution of naked singularities,''
  JHEP {\bf 0008}, 052 (2000)
  [arXiv:hep-th/0007191].


%

%

\bibitem{Karch:2002sh}
A.~Karch and E.~Katz,
``Adding flavor to AdS/CFT,''
JHEP {\bf 0206}, 043 (2002)
[arXiv:hep-th/0205236].


\bibitem{Kruczenski:2003be}
  M.~Kruczenski, D.~Mateos, R.~C.~Myers and D.~J.~Winters,
  JHEP {\bf 0307}, 049 (2003)
  [arXiv:hep-th/0304032].
  M.~Kruczenski, D.~Mateos, R.~C.~Myers and D.~J.~Winters,
  JHEP {\bf 0405}, 041 (2004)
  [arXiv:hep-th/0311270].
  J.~Babington, J.~Erdmenger, N.~J.~Evans, Z.~Guralnik and I.~Kirsch,
  Phys.\ Rev.\  D {\bf 69}, 066007 (2004)
  [arXiv:hep-th/0306018].
  D.~Mateos, R.~C.~Myers and R.~M.~Thomson,
  Phys.\ Rev.\ Lett.\  {\bf 97}, 091601 (2006)
  [arXiv:hep-th/0605046].
  T.~Albash, V.~G.~Filev, C.~V.~Johnson and A.~Kundu,
  arXiv:hep-th/0605088.
  T.~Sakai and S.~Sugimoto,
  Prog.\ Theor.\ Phys.\  {\bf 113}, 843 (2005)
  [arXiv:hep-th/0412141].

\bibitem{Casero:2006pt}
R.~Casero, C.~Nunez and A.~Paredes,
``Towards the string dual of N = 1 SQCD-like theories,''
Phys.\ Rev.\  D {\bf 73}, 086005 (2006)
[arXiv:hep-th/0602027].

\bibitem{Benini:2006hh}
F.~Benini, F.~Canoura, S.~Cremonesi, C.~Nunez and A.~V.~Ramallo,
``Unquenched flavors in the Klebanov-Witten model,''
JHEP {\bf 0702}, 090 (2007)
[arXiv:hep-th/0612118].

\bibitem{Benini:2007gx}
F.~Benini, F.~Canoura, S.~Cremonesi, C.~Nunez and A.~V.~Ramallo,
``Backreacting Flavors in the Klebanov-Strassler Background,''
arXiv:0706.1238 [hep-th].

\bibitem{Nunez:2003cf}
C.~Nunez, A.~Paredes and A.~V.~Ramallo,
``Flavoring the gravity dual of N = 1 Yang-Mills with probes,''
JHEP {\bf 0312}, 024 (2003)
[arXiv:hep-th/0311201].

\bibitem{Papadopoulos:2000gj}
  G.~Papadopoulos and A.~A.~Tseytlin,
  ``Complex geometry of conifolds and 5-brane wrapped on 2-sphere,''
  Class.\ Quant.\ Grav.\  {\bf 18}, 1333 (2001)
  [arXiv:hep-th/0012034].

\bibitem{Gubser:2001eg}
  S.~S.~Gubser, A.~A.~Tseytlin and M.~S.~Volkov,
  ``Non-Abelian 4-d black holes, wrapped 5-branes, and their dual
  descriptions,''
  JHEP {\bf 0109}, 017 (2001)
  [arXiv:hep-th/0108205].

 \bibitem{Seiberg:1997zk}
 N.~Seiberg,
 ``New theories in six dimensions and matrix description of M-theory on  T**5 and T**5/Z(2),''
 Phys.\ Lett.\  B {\bf 408}, 98 (1997)
 [arXiv:hep-th/9705221].
 
\bibitem{Losev:1997hx}
  A.~Losev, G.~W.~Moore and S.~L.~Shatashvili,
  ``M \& m's,''
  Nucl.\ Phys.\  B {\bf 522}, 105 (1998)
  [arXiv:hep-th/9707250].

 \bibitem{Aharony:1999ks}
 O.~Aharony,
 ``A brief review of 'little string theories',''
 Class.\ Quant.\ Grav.\  {\bf 17}, 929 (2000)
 [arXiv:hep-th/9911147].

\bibitem{Itzhaki:1998dd}
  N.~Itzhaki, J.~M.~Maldacena, J.~Sonnenschein and S.~Yankielowicz,
  ``Supergravity and the large N limit of theories with sixteen
  supercharges,''
  Phys.\ Rev.\  D {\bf 58}, 046004 (1998)
  [arXiv:hep-th/9802042].

\bibitem{Gauntlett:2001ps}
  J.~P.~Gauntlett, N.~Kim, D.~Martelli and D.~Waldram,
  ``Wrapped fivebranes and N = 2 super Yang-Mills theory,''
  Phys.\ Rev.\  D {\bf 64}, 106008 (2001)
  [arXiv:hep-th/0106117].

\bibitem{Bershadsky:1995qy}
  M.~Bershadsky, C.~Vafa and V.~Sadov,
  ``D-Branes and Topological Field Theories,''
  Nucl.\ Phys.\  B {\bf 463}, 420 (1996)
  [arXiv:hep-th/9511222].

\bibitem{Gubser:2000nd}
S.~S.~Gubser,
``Curvature singularities: The good, the bad, and the naked,''
Adv.\ Theor.\ Math.\ Phys.\  {\bf 4}, 679 (2002)
[arXiv:hep-th/0002160].

\bibitem{Casero:2007jj}
R.~Casero, C.~Nunez and A.~Paredes,
``Elaborations on the String Dual to N=1 SQCD,''
arXiv:0709.3421 [hep-th].

\bibitem{Bertoldi:2007sf}
G.~Bertoldi, F.~Bigazzi, A.~L.~Cotrone and J.~D.~Edelstein,
``Holography and Unquenched Quark-Gluon Plasmas,''
arXiv:hep-th/0702225.

\bibitem{Bigazzi:2005md}
F.~Bigazzi, R.~Casero, A.~L.~Cotrone, E.~Kiritsis and A.~Paredes,
``Non-critical holography and four-dimensional CFT's with fundamentals,''
JHEP {\bf 0510}, 012 (2005)
[arXiv:hep-th/0505140].

\bibitem{Karch:2000gx}
  A.~Karch and L.~Randall,
  ``Open and closed string interpretation of SUSY CFT's on branes with
  boundaries,''
  JHEP {\bf 0106}, 063 (2001)
  [arXiv:hep-th/0105132].

\bibitem{Bachas:2000fr}
  C.~Bachas and M.~Petropoulos,
  ``Anti-de-Sitter D-branes,''
  JHEP {\bf 0102}, 025 (2001)
  [arXiv:hep-th/0012234].

\bibitem{Bergshoeff:1997kr}
  E.~Bergshoeff, R.~Kallosh, T.~Ortin and G.~Papadopoulos,
  ``kappa-symmetry, supersymmetry and intersecting branes,''
  Nucl.\ Phys.\  B {\bf 502}, 149 (1997)
  [arXiv:hep-th/9705040].

\bibitem{Candelas:1989js}
P.~Candelas and X.~C.~de la Ossa,
``Comments on Conifolds,''
Nucl.\ Phys.\  B {\bf 342}, 246 (1990).

\bibitem{Maldacena:2000mw}
J.~M.~Maldacena and C.~Nunez,
``Supergravity description of field theories on curved manifolds and a no go theorem,''
Int.\ J.\ Mod.\ Phys.\  A {\bf 16}, 822 (2001)
[arXiv:hep-th/0007018].

\bibitem{Klebanov:2000nc}
I.~R.~Klebanov and A.~A.~Tseytlin,
``Gravity duals of supersymmetric SU(N) x SU(N+M) gauge theories,''
Nucl.\ Phys.\  B {\bf 578}, 123 (2000)
[arXiv:hep-th/0002159].

\bibitem{Maldacena:1998im}
J.~M.~Maldacena,
``Wilson loops in large N field theories,''
Phys.\ Rev.\ Lett.\  {\bf 80}, 4859 (1998)
[arXiv:hep-th/9803002].

\bibitem{Rey:1998ik}
S.~J.~Rey and J.~T.~Yee,
``Macroscopic strings as heavy quarks in large N gauge theory and  anti-de Sitter supergravity,''
Eur.\ Phys.\ J.\  C {\bf 22}, 379 (2001)
[arXiv:hep-th/9803001].

\bibitem{Peet:1998wn}
A.~W.~Peet and J.~Polchinski,
``UV/IR relations in AdS dynamics,''
Phys.\ Rev.\  D {\bf 59}, 065011 (1999)
[arXiv:hep-th/9809022].

\bibitem{Caceres:2005yx}
  E.~Caceres and C.~Nunez,
  ``Glueballs of super Yang-Mills from wrapped branes,''
  JHEP {\bf 0509}, 027 (2005)
  [arXiv:hep-th/0506051].
\bibitem{Policastro:2001yc}
  G.~Policastro, D.~T.~Son and A.~O.~Starinets,
  ``The shear viscosity of strongly coupled N = 4 supersymmetric Yang-Mills
  plasma,''
  Phys.\ Rev.\ Lett.\  {\bf 87}, 081601 (2001)
  [arXiv:hep-th/0104066].
\bibitem{Policastro:2002se}
  G.~Policastro, D.~T.~Son and A.~O.~Starinets,
  ``From AdS/CFT correspondence to hydrodynamics,''
  JHEP {\bf 0209}, 043 (2002)
  [arXiv:hep-th/0205052].
\bibitem{Kovtun:2004de}
  P.~Kovtun, D.~T.~Son and A.~O.~Starinets,
  ``Viscosity in strongly interacting quantum field theories from black hole
  physics,''
  Phys.\ Rev.\ Lett.\  {\bf 94}, 111601 (2005)
  [arXiv:hep-th/0405231].


\bibitem{Herzog:2006gh}
  C.~P.~Herzog, A.~Karch, P.~Kovtun, C.~Kozcaz and L.~G.~Yaffe,
  JHEP {\bf 0607}, 013 (2006)
  [arXiv:hep-th/0605158].
  S.~S.~Gubser,
  Phys.\ Rev.\  D {\bf 74}, 126005 (2006)
  [arXiv:hep-th/0605182].
  H.~Liu, K.~Rajagopal and U.~A.~Wiedemann,
  Phys.\ Rev.\ Lett.\  {\bf 98}, 182301 (2007)
  [arXiv:hep-ph/0607062].
  H.~Liu, K.~Rajagopal and U.~A.~Wiedemann,
  Phys.\ Rev.\ Lett.\  {\bf 97}, 182301 (2006)
  [arXiv:hep-ph/0605178].
  E.~Caceres and A.~Guijosa,
  JHEP {\bf 0612}, 068 (2006)
  [arXiv:hep-th/0606134].
  E.~Caceres and A.~Guijosa,
  JHEP {\bf 0611}, 077 (2006)
  [arXiv:hep-th/0605235].
  A.~Buchel,
  Phys.\ Rev.\  D {\bf 74}, 046006 (2006)
  [arXiv:hep-th/0605178].
  E.~Caceres, M.~Natsuume and T.~Okamura,
  JHEP {\bf 0610}, 011 (2006)
  [arXiv:hep-th/0607233].
  M.~Chernicoff and A.~Guijosa,
  JHEP {\bf 0702}, 084 (2007)
  [arXiv:hep-th/0611155].
  S.~D.~Avramis and K.~Sfetsos,
  JHEP {\bf 0701}, 065 (2007)
  [arXiv:hep-th/0606190].
  J.~Mas,
  JHEP {\bf 0603}, 016 (2006)
  [arXiv:hep-th/0601144].
  N.~Armesto, J.~D.~Edelstein and J.~Mas,
  JHEP {\bf 0609}, 039 (2006)
  [arXiv:hep-ph/0606245].
 S.~D.~Avramis, K.~Sfetsos and D.~Zoakos,
  Phys.\ Rev.\  D {\bf 75}, 025009 (2007)
  [arXiv:hep-th/0609079].
  M.~Chernicoff, J.~A.~Garcia and A.~Guijosa,
  JHEP {\bf 0609}, 068 (2006)
  [arXiv:hep-th/0607089].
  M.~Natsuume and T.~Okamura,
  JHEP {\bf 0709}, 039 (2007)
  [arXiv:0706.0086 [hep-th]].







\bibitem{Aharony:2002vp}
  O.~Aharony, E.~Schreiber and J.~Sonnenschein,
  JHEP {\bf 0204}, 011 (2002)
  [arXiv:hep-th/0201224].
  V.~Borokhov and S.~S.~Gubser,
  JHEP {\bf 0305}, 034 (2003)
  [arXiv:hep-th/0206098].
  N.~J.~Evans, M.~Petrini and A.~Zaffaroni,
  JHEP {\bf 0206}, 004 (2002)
  [arXiv:hep-th/0203203].


\bibitem{Bergshoeff:1996tu}
  E.~Bergshoeff and P.~K.~Townsend,
  ``Super D-branes,''
  Nucl.\ Phys.\  B {\bf 490}, 145 (1997)
  [arXiv:hep-th/9611173].

\bibitem{Bergshoeff:1998ha}
  E.~Bergshoeff and P.~K.~Townsend,
  ``Super-D branes revisited,''
  Nucl.\ Phys.\  B {\bf 531}, 226 (1998)
  [arXiv:hep-th/9804011].
\bibitem{Schwarz:1983qr}
J.~H.~Schwarz, ``Covariant Field Equations Of Chiral N=2 D=10 Supergravity,''
Nucl.\ Phys.\  B {\bf 226}, 269 (1983).

\end{thebibliography}
\end{document}